\documentclass[useAMS,usenatbib]{mn2e}

\usepackage{graphicx}
\usepackage{times}%
\usepackage{natbib}
\usepackage{graphics}
\usepackage{epsfig}
\usepackage{array}
\usepackage{stfloats}
\usepackage{fixltx2e}
\usepackage{amsmath}

\newcommand{\ie}{{i.e.}}
\newcommand{\eg}{{e.g.}}
\newcommand{\gsim}{\,\lower2truept\hbox{${>\atop\hbox{\raise4truept\hbox{$\sim$}}}$}\,}

\newcommand{\vv}{~~~,}

\def\eg{{\rm e.g.$\,$}}
\def\ie{{\rm i.e.$\,$}}

\newcommand{\be}{\begin{equation}}
\newcommand{\ee}{\end{equation}}
\newcommand{\bea}{\begin{eqnarray}}
\newcommand{\eea}{\end{eqnarray}}

\renewcommand{\vec}[1]{ {\bmath #1} } 

\def\ltsima{$\; \buildrel < \over \sim \;$}
\def\simlt{\lower.5ex\hbox{\ltsima}}
\def\gtsima{$\; \buildrel > \over \sim \;$}
\def\simgt{\lower.5ex\hbox{\gtsima}}

\title[Interacting Dark Energy and Structure Formation]{Clarifying the effects of interacting dark energy on linear and nonlinear structure formation processes}

\author[M. Baldi]{Marco Baldi
\\Excellence Cluster Universe, Boltzmannstr.~2, D-85748 Garching, Germany
\\University Observatory, Ludwig-Maximillians University Munich, Scheinerstr. 1, D-81679 Munich, Germany
\\marco.baldi@universe-cluster.de}

\begin{document}


\pagerange{\pageref{firstpage}--\pageref{lastpage}} \pubyear{2010}

\maketitle

\label{firstpage}

\begin{abstract}	

We present a detailed numerical study of the impact that cosmological models featuring a direct interaction between
the Dark Energy component that drives the accelerated expansion of the Universe and Cold Dark Matter can have on the linear
and nonlinear stages of structure formation. By means of a series of collisionless N-body simulations we study the influence that
each of the different effects characterizing these cosmological models -- which include among others a fifth force, a time variation of
particle masses, and a velocity-dependent acceleration -- separately have on the growth of density perturbations
and on a series of observable quantities related to linear and nonlinear cosmic structures, as the matter power spectrum, the gravitational
bias between baryons and Cold Dark Matter, the halo mass function and the halo density profiles. We perform our analysis applying
and comparing different numerical approaches previously adopted in the literature, and we address the partial discrepancies recently
claimed in a similar study by \citet{Li_Barrow_2010b} with respect to the first outcomes of \citet{Baldi_etal_2010}, which are found to be 
related to the specific numerical approach adopted in the former work. 
Our results fully confirm the conclusions of \citet{Baldi_etal_2010} and show that when linear and nonlinear effects of the interaction
between Dark Energy and Cold Dark Matter are properly disentangled, the velocity-dependent acceleration is the leading
effect acting at nonlinear scales, and in particular is the most important mechanism in lowering the concentration of Cold Dark Matter halos.

\end{abstract}
 


\section{Introduction}
\label{i}

The two pillars on which the presently accepted standard cosmological model is based,
as witnessed by its acronym $\Lambda $CDM, are a cosmological constant $\Lambda $
which determines an acceleration of the expansion of the Universe, and a new type of non relativistic massive
particles -- which go under the name of Cold Dark Matter (CDM) -- that source the gravitational potential
wells in which cosmic structures can form. 
Far from being the ultimate description of our Universe, the $\Lambda $CDM model represents an
efficient way to parametrize our ignorance about the fundamental constituents of roughly 95\% of its
energy content.

Although the combination of a cosmological constant $\Lambda $ and of a significant fraction
of CDM particles provides a very good fit to most of the presently available observations \citep[see \eg][]{wmap5,wmap7,Percival_etal_2001,Cole_etal_2005,Reid_etal_2010,Riess_etal_1998,Perlmutter_etal_1999,SNLS,Kowalski_etal_2008,Percival_etal_2009}, 
the fundamental nature 
of the two main components of the Universe remains an open question.
On one side, in fact, the observed value of the energy scale associated with the cosmological constant differs by tens of orders of magnitude
from its theoretical predictions, thereby making of the cosmological constant an extremely fine-tuned parameter of the model.
On the other side, all the proposed CDM candidates have so far evaded any attempt of direct detection and their 
properties can still be inferred only by cosmological and astrophysical observations.

It is in this context that alternative models have been proposed, by taking the standard $\Lambda $CDM as
an asymptotic state of more complex underlying scenarios. These generally involve, as a first step, the promotion
of the cosmological constant $\Lambda $ to a dynamical quantity -- dubbed dark energy (DE) -- which evolves during the expansion history of the Universe
just as all its other physical constituents. A convenient way to represent this scenario involves the dynamical
evolution of a classical scalar field in a self interaction potential, which goes under the name of quintessence
\citep{Wetterich_1988,Ratra_Peebles_1988} or k-essence \citep{kessence}. 

Other possible alternatives with respect to the standard cosmological scenario are given by modifications
of the laws of gravity at large scales or at low spatial curvatures \citep[\eg][]{Starobinsky_1980,Hu_Sawicki_2007}, or by relaxing the assumptions of homogeneity and isotropy of the 
Universe \citep[see \eg][]{GarciaBellido_Haugboelle_2008}.

Particular attention has been devoted, in recent years, to the idea that a dynamical DE scalar field might have direct
interactions (besides gravity) with other cosmic fluids, by directly exchanging energy-momentum \citep{Wetterich_1995,Amendola_2000,Farrar2004,Amendola_Baldi_Wetterich_2008}. 
While such direct interactions
with standard model particles would be heavily restricted by available observational constraints, the same does not happen
for the case of a selective interaction with CDM particles only, as first suggested by 
\citet{Damour_Gibbons_Gundlach_1990}, for which observational bounds are much weaker.

These models have been extensively studied in the recent past, with a particular focus on their possible distinctive effects on structure formation.
In fact, all interacting DE models predict the existence of an additional attractive force -- mediated by the DE scalar field -- between massive particles
which is expected to alter the growth of density perturbations.
Besides this ``fifth-force", coupled massive particles also experience, in interacting DE models, a ``modified inertia" determined by the energy-momentum transfer between
the DE and the CDM sectors.

The combination of these new physical effects and of the modified cosmic expansion -- that arises as a consequence of the dynamical nature of the DE field -- determines a 
modified evolution of density perturbations thereby providing possible observational signatures of the models.
The effects of interacting DE models on the evolution of cosmic structures have been studied both in the linear regime \citep{Amendola_2000,Amendola_2004,Pettorino_Baccigalupi_2008,DiPorto_Amendola_2008,Baldi_2010} and in the nonlinear regime by relying on
simplified configurations, as for the case of spherical collapse \citep{Mainini_Bonometto_2006,Wintergerst_Pettorino_2010}, or in full generality by means of N-body simulations \citep[\eg by][]{Baldi_etal_2010,Baldi_2010,Li_Barrow_2010}.

In this work we present an extension of previous analyses about how the different new physical effects featured by interacting 
DE models individually contribute to the overall modifications of the properties of linear and nonlinear cosmic structures, and
we try to clarify some confusion that has been made in the literature concerning the relative importance of the different effects in these
two different regimes.

The paper is organized as follows: in Sec.~\ref{cde} we briefly review the main equations of interacting DE models both concerning
the background evolution and the growth of linear perturbations, with a particular focus on the issue of model normalization which will be central
for the rest of the present study; in Sec.~\ref{mog} we discuss the different effects that contribute to a 
modification of newtonian dynamics at small scales for CDM particles; 
in Sec.~\ref{sim} we present our set of simulations, describing which effects have been included in 
each simulation, and how the relevant quantities have been normalized in order to allow a meaningful comparison of the different outcomes;
in Sec.~\ref{res} we present our results concerning linear and nonlinear properties of cosmic structures under different implementations
of the interacting DE effects; finally, in Sec.~\ref{concl} we draw our conclusions.

\section{Interacting scalar fields} \label{cde}

We consider interacting DE models where the role of DE is played by a scalar field $\phi $ evolving in 
a self-interaction potential $V(\phi )$, and where the interaction with CDM particles is represented by 
a source term in the respective continuity equations for the two fluids\footnote{Throughout the paper we will use units in which 
the reduced Planck mass is assumed to be unity, $M_{Pl}\equiv 1/\sqrt{8\pi G} = 1$.}:
\begin{eqnarray}
\label{cons_c} \dot{\rho }_{c} +3H \rho_{c} &=& - \beta (\phi )\dot{\phi } \rho_{c} \\
\label{cons_phi} \dot{\rho }_{\phi} +3H \rho_{\phi} &=& + \beta (\phi )\dot{\phi } \rho_{c} \vv 
\end{eqnarray}
where $\rho _{c}$ is the CDM density, $\rho _{\phi }\equiv \dot{\phi }^{2}/2 + V(\phi )$ is the DE density, $H$ is the Hubble
function, and where an overdot represents a derivative with respect to cosmic time $t$. 
The function $\beta (\phi )$ determines the strength of the DE-CDM interaction and together with the 
scalar potential $V(\phi )$ fully specifies the model.

Interacting DE models described by Eqs.~(\ref{cons_c},\ref{cons_phi}), where all the other cosmic fluids (as radiation,
baryonic matter, massive neutrinos, etc...) obey standard continuity equations, have been widely studied in the past
both for the case of a constant coupling \citep[as \eg by][and references therein]{Wetterich_1995,Amendola_2000,Amendola_2004,Pettorino_Baccigalupi_2008}
and for the more general situation of a field depentend coupling \citep{Baldi_2010}. In the present work we will focus
on the former situation, and we will therefore assume $\beta (\phi ) = \beta $ for the rest of the paper.
We will also always assume an exponential form for the potential $V(\phi )\propto e^{-\alpha \phi}$,
with $\alpha =0.1$.

The background evolution of constant coupling models is characterized by a scaling regime during matter domination
where the two interacting fluids (DE and CDM in our case) share a constant ratio of the total energy budget of the universe\footnote{We always assume spatial flatness
such that $\rho _{{\rm tot}} = 3H^{2}$.}
 (known as the $\phi $MDE regime), 
thereby determining the presence of a sizeable fraction of Early DE (EDE hereafter) which alters the standard
expansion history with respect to a corresponding $\Lambda $CDM model with the same set of cosmolgical parameters
at the present time. This scaling regime is sustained (for the case of positive couplings on which we focus in the present work) 
by the energy transfer from the CDM particles to the DE scalar field, which determines in turn a decrease in time of the mass
of CDM particles, according to the modified continuity equation (\ref{cons_c}).
Therefore, a modified expansion rate at $z > 0$ and a time evolution of CDM particle masses are two common features of any interacting DE
model which can affect the growth rate of density perturbations. 

For the latter effect, it is clearly important the normalization of the CDM masses:
if one wants to compare models that share the same cosmological parameters at $z=0$, then the mass of CDM particles within 
interacting DE models will be larger in the past,
corresponding to an effectively higher value of the CDM density $\rho _{c}$.
However, it is important to stress here the distinction between the mass variation, \ie the fact that $\dot{m_{c}}\neq 0$, and the mass normalization,
\ie the effective $\rho _{c}(z)$ during the expansion history of the universe. This distinction is particularly important for what we will show in the rest 
of the paper, since the former effect is found to have significant implications for the nonlinear regime of structure formation and for the internal
dynamics of collapsed objects, while the latter is primarily affecting the linear evolution of density perturbations. Not recognizing this
distinction can cause some confusion in the study of the nonlinear effects of interacting DE models, as we will discuss below.

Besides these two features, the study of the linear perturbations evolution in interacting DE models \citep{Amendola_2000,Amendola_2004,Pettorino_Baccigalupi_2008} has led to the identification of other two relevant effects
which directly influence the processes of structure evolution. The dynamic equation for CDM density perturbations in interacting DE scenarios
\begin{equation}
\label{linear_evolution}
\ddot{\delta _{c}} + \left( 2H - \beta \dot{\phi }\right)\dot{\delta _{c}} - \frac{3}{2}H^2\left[\left( 1+2\beta ^{2}\right) \Omega _{c}\delta _{c} + \Omega _{b}\delta _{b}\right] = 0\,,
\end{equation}
shows in fact also the presence of an additional friction term directly proportional to the coupling
\begin{equation}
\label{friction_term}
-\beta\dot{\phi }\dot{\delta _{c}} \,,
\end{equation}
and of an effective enhancement of the gravitational pull for CDM fluctuations by a factor $(1 + 2\beta ^{2})$ which is known as the
``fifth-force". Both the extra friction term and the fifth-force accelerate the growth of CDM density perturbations in the linear regime
as clearly shown by Eqn.~(\ref{linear_evolution}).

\subsection{Model normalization}

It is of particular relevance for the kind of analysis carried out in the present work
to clarify which normalization is assumed for the different cosmological models under investigation
(that in this case are a $\Lambda $CDM cosmology and an interacting DE scenario with constant coupling $\beta $) 
and for the different
numerical realizations of the latter in the test simulations that we are going to use for our discussion.
The normalization of the models concerns both the background and the linear perturbations evolution in the different 
test cases, which we separately discuss here.

\subsubsection{Background normalization}
\label{norm_bckgd}

\begin{figure*}
\begin{center}
\includegraphics[scale=0.4]{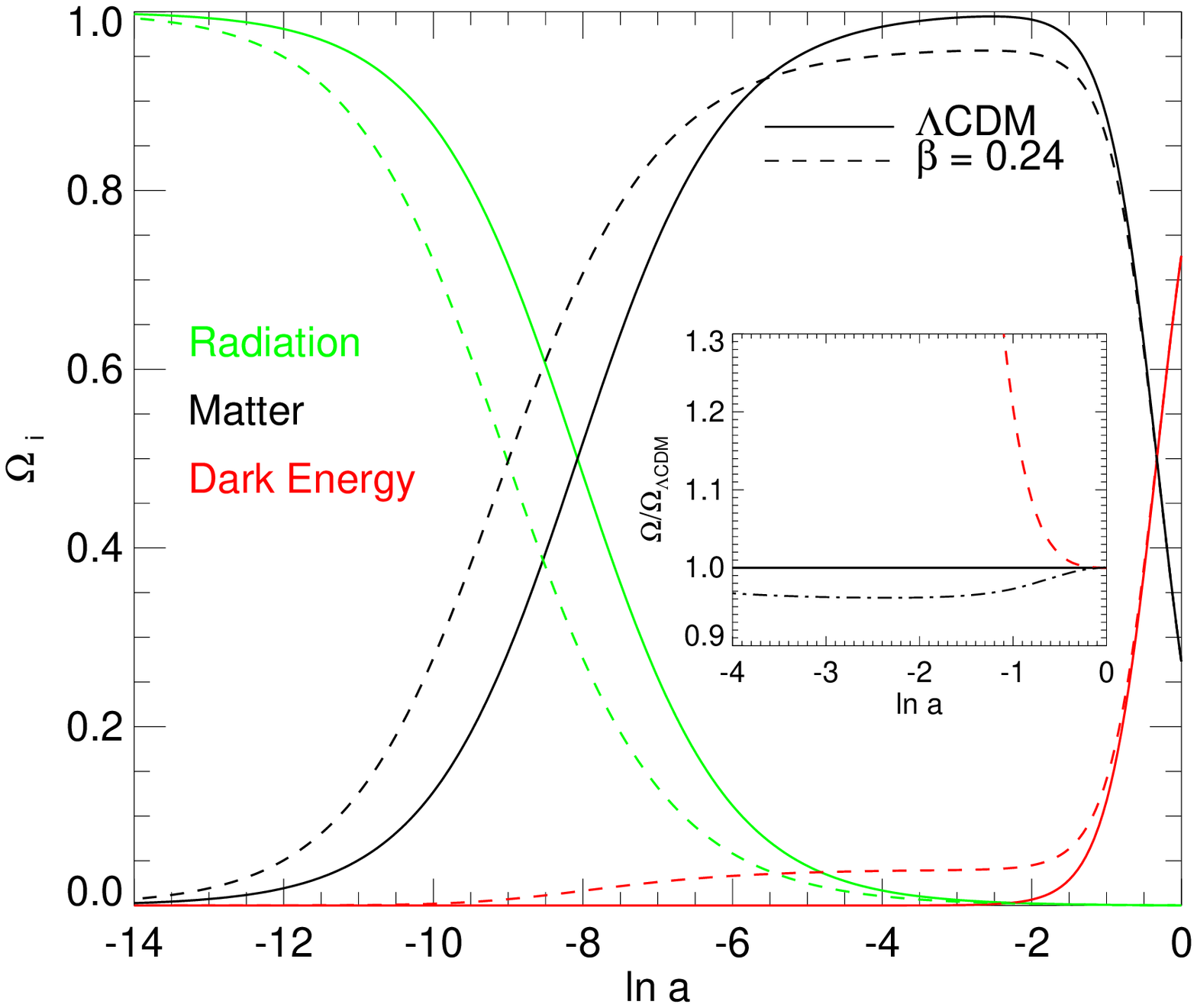}
\includegraphics[scale=0.4]{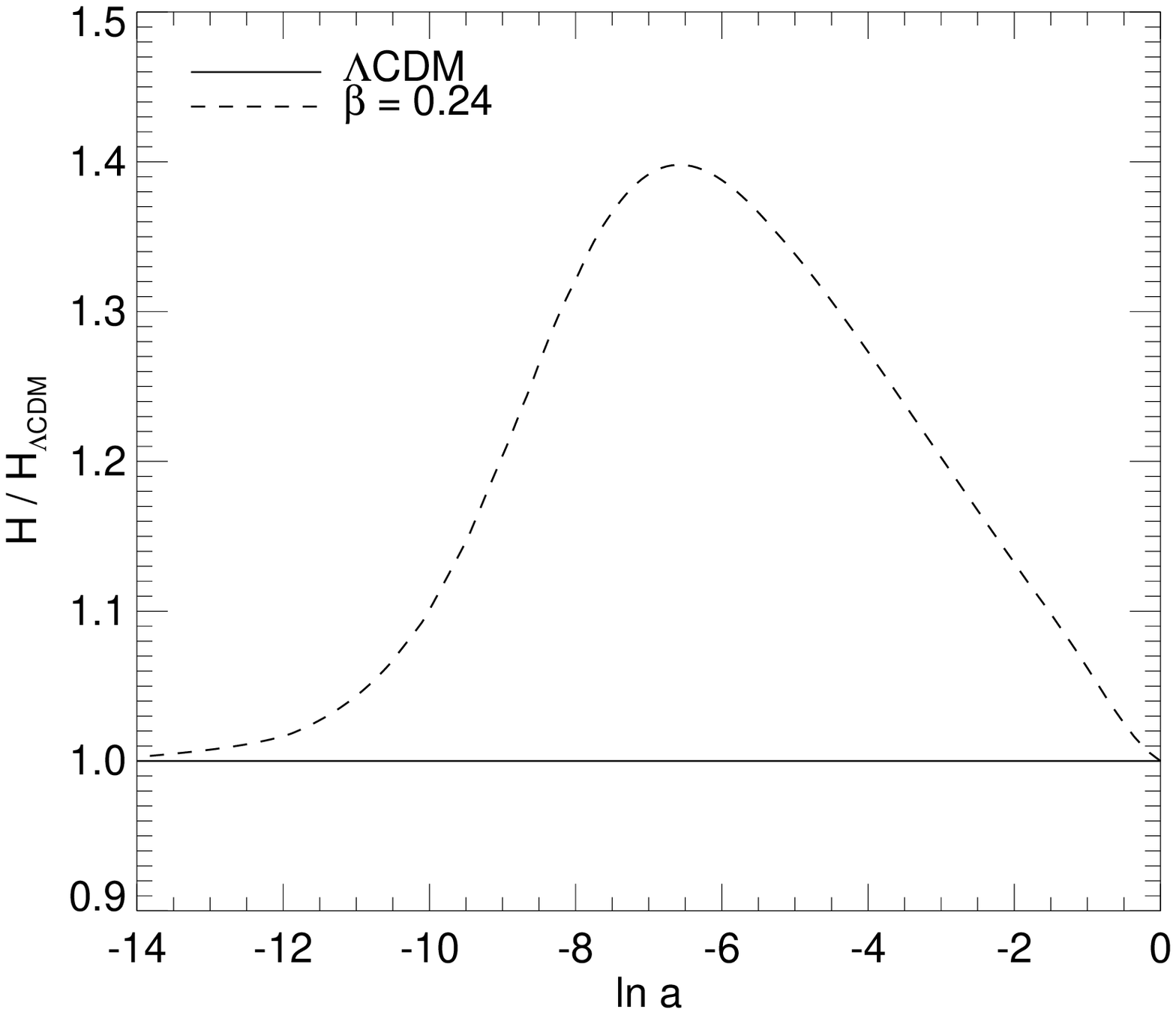}
  \caption{(Color online) {\em Left panel}: The background evolution of a standard $\Lambda $CDM model and of an interacting DE model with $\beta = 0.24$. In this example
  we have not included the uncoupled fraction of baryonic matter to simplify the plot. The figure shows the evolution of the dimensionless density in radiation (green), matter (black) and dark energy (red) as a function of the e-folding time (defined as the logarithm of the scale factor $a$) for $\Lambda $CDM (solid) and interacting DE (dashed). The smaller plot shows the ratio of the same quantities to the $\Lambda $CDM case. {\em Right panel}: The ratio of the Hubble function of the same two models with respect to $\Lambda $CDM as a function of the e-folding time. A larger Hubble function with respect to $\Lambda $CDM is consistent with all previous studies on interacting DE and EDE, as explained in the text.}
\label{fig:background}
\end{center}
\end{figure*}
\normalsize
In the present work we adopt the same type of normalization assumed in the previous works of \citet{Baldi_etal_2010} and \citet{Baldi_2010}
for the background evolution, where both the $\Lambda $CDM and the interacting DE model share the same 
background cosmological parameters at $z=0$. This is realized by assuming a specific set of cosmological parameters (given by
$H_{0}, \Omega_{c}, \Omega_{b}, \Omega _{r}, \Omega _{{\rm DE}}$) in accordance with the latest observational constraints from the
WMAP satellite \citep{wmap7} and by then integrating backwards in time the full system of background equations for each specific model.
As a consequence of this procedure, both the $\Lambda $CDM and the interacting DE model will have exactly the same cosmological parameters at $z=0$
and will therefore differ from each other at higher redshifts. In particular, the mass of CDM particles will be larger at high redshifts due to the interaction
with the DE scalar filed $\phi $ and to the continuous flow of energy from the former to the latter. Due to the existence of the $\phi $MDE scaling
regime, where a sizeable fraction of EDE is present during matter domination, and to the requirement of spatial flatness 
($\Omega _{c} + \Omega _{b} + \Omega _{r} + \Omega _{{\rm DE}} = 1$), if a larger fraction of DE is present in interacting DE models during matter domination,
then the total value of $\Omega _{c}$ will necessarily have to be smaller at the same time as compared to $\Lambda $CDM. A larger value of
$\rho _{c}$ and a smaller value of $\Omega _{c} \equiv \rho _{c}/3H^{2}$ then obviously implies a larger value of the Hubble function during $\phi $MDE
with respect to $\Lambda $CDM. This is what is shown in Fig.~\ref{fig:background}, where the evolution of the cosmological density parameters and of
the Hubble function are shown for $\Lambda $CDM and for our interacting DE model. This is consistent with most of the other studies on interacting DE and EDE models
\citep[as \eg][]{Doran_Schwindt_Wetterich_2001,Wetterich_2004,Pettorino_Baccigalupi_2008,Francis_Lewis_Linder_2008,Grossi_2008}.

This type of normalization therefore allows to compare different models that have exactly the same cosmological parameters at $z=0$,
thereby providing a homogeneous set of models to be confronted with each other. 
Other studies have adopted different types of normalization \citep[as \eg in ][]{Li_Barrow_2010,Li_Barrow_2010b} where
different interacting DE models are normalized by assuming the same initial conditions in the early universe for the background dynamical quantities
and by evolving these initial conditions forward in time with a trial and error procedure until suitable solutions are found.
This procedure, however, clearly does not guarantee to get the same values for the background cosmological parameters at $z=0$.

\subsubsection{Linear perturbations normalization}

A completely independent choice of normalization between different interacting DE models concerns the amplitude of linear density perturbations.
As a consequence of the new physical processes and of the different background evolution, in fact, interacting DE models will have a significantly
different growth rate of density perturbations as compared to $\Lambda $CDM, where the deviation will depend on the strength of the coupling 
\citep[see][for a detailed description of the growth factor in interacting DE scenarios]{Baldi_etal_2010,DiPorto_Amendola_2008}.
When setting up the initial conditions for N-body simulations, one is therefore allowed to choose whether to set the desired value of the linear density 
perturbations at high redshifts, by fixing the amplitude according to 
the observed value of the scalar amplitude ${\cal A}_{s}$ from CMB data, or at the present time, by assuming the present value of $\sigma _{8}$ as determined by low-redshift measurements as normalizations.
In the former case, different models will start with identical density amplitudes at early times and will end up with different values of $\sigma _{8}$ at $z=0$ \citep[as shown \eg in][]{Baldi_Pettorino_2010},
while in the latter the amplitude of the initial density fluctuations will have to be scaled with the appropriate growth factor of each specific model in order
to give the same value of $\sigma _{8}$ at present \citep[as done in ][]{Baldi_etal_2010,Baldi_2010,Li_Barrow_2010}. 

Both types of normalization have been used in the past in different numerical studies aimed at different types of analysis of nonlinear structure formation
within interacting DE models. However, the normalization at $z=0$ is clearly the preferred choice if one aims at directly comparing the effects of interacting DE on the nonlinear
dynamics of CDM particles within collapsed structures, since with such normalization all the models will share at the present time the same cosmological
parameters and the same linear density perturbations normalization, and all the differences in the properties of low redshift structures
will be related to the nonlinear behavior only.

It is nevertheless important to stress here that if any of the new physical effects of interacting DE  is artificially suppressed in an N-body run,
then the effective linear growth factor in the simulation will change and a different value of $\sigma _{8}$ at $z=0$ will be obtained even if the initial conditions were normalized 
with the preferred procedure described above. This might provide interesting information on the impact that the specific effect which has been suppressed has on the linear regime of structure formation, but could also introduce some confusion in the study of the nonlinear regime, since nonlinear structures forming in such test simulation will be embedded in a linear
density field with a different normalization.

\section{Modified gravitational dynamics}
\label{mog}

If we want to follow the evolution of density perturbations beyond the linear regime and be able to predict the features that
interacting DE imprints on the highly nonlinear objects that we can directly observe in the sky, Eqn.~(\ref{linear_evolution})
is no longer sufficient and we need to rely on numerical integrations. In order to do so, one has to identify how the 
interaction between DE and CDM affects the laws of newtonian dynamics that govern the evolution of structure formation
in the newtonian limit of General Relativity and implement these effects into N-body algorithms.

\citet{Baldi_etal_2010} have shown that the acceleration equation for a CDM
particle in an interacting DE cosmology for the case of a light scalar field (\ie a scalar field model for which $m_{\phi }\equiv {\rm d}^{2}V/{\rm d}\phi ^{2} \ll H$) takes the form \citep[see][for the details of the derivation]{Baldi_etal_2010}:
\begin{equation}
\label{acceleration_equation}
\dot{\vec{v}}_{i} = \beta \dot{\phi } \vec{v}_{i} + \sum_{j \ne i}\frac{G(1+2\beta ^{2})m_{j}\vec{r}_{ij}}{|\vec{r}_{ij}|^{3}} \,,
\end{equation}
where $\vec{v}_{i}$ is the velocity of the $i$-th particle, $\vec{r}_{ij}$ is the vector distance between the $i$-th and the $j$-th particles, and the sum extends to all the CDM particles in the universe.

By having a look at Eqn.~(\ref{acceleration_equation}), one can clearly identify the same coupling-dependent terms
already encountered for the linear perturbations equation (\ref{linear_evolution}). The friction term of Eq.~(\ref{friction_term})
now appears as a ``velocity-dependent" acceleration
\begin{equation}
\vec{a}_{v} = \beta \dot{\phi }\vec{v}
\label{velocity_acc}
\end{equation}
which depends on the vectorial velocity of CDM particles, while the ``fifth-force"
term appears in the same form as in Eq.~(\ref{linear_evolution}), \ie as an effective rescaling of the gravitational constant $G$ between CDM particles pairs by a factor $(1+2\beta^{2})$.

It is however very important to notice a striking difference between the linear and the nonlinear regimes, that will have significant implications
for the discussion carried out in this paper: while in the linear regime the friction term always accelerates the growth of structures, in the nonlinear
case this depends on the relative orientations of the velocity and of the gravitational acceleration of each CDM particle. As a consequence, 
a particle moving towards the local potential minimum will experience an effectively larger potential gradient while a particle moving 
away from the local potential minimum will conversely feel an effectively smaller potential gradient. For the realistic situation of nonlinear virialized objects, 
where tangential velocities are non negligible with respect to radial velocities, the velocity-dependent acceleration will therefore have a completely different 
effect than in the linear regime.

For this reason, when comparing the properties of nonlinear structures one should avoid to consider the linear friction term ad the nonlinear velocity-dependent acceleration
as a single phenomenon, and always distinguish between its linear and nonlinear behavior. As we will show below, not making this 
clear distinction between the two regimes can cause some further confusion in the determination of which are the most relevant effects of interacting DE
for the nonlinear dynamics of CDM particles.

\section{Simulations}
\label{sim}

The study of the nonlinear effects of interacting DE models has been approached in recent years by means of suitably modified N-body
algorithms. The first hydrodynamical high-resolution N-body simulations of interacting DE models have been performed with a modification
of the parallel TreePM code {\small GADGET-2} \citep{gadget-2} and presented in \citet{Baldi_etal_2010}.
Other studies have then been carried out by means of mesh or Tree based N-body algorithms, but without
hydrodynamics \citep[see \eg][]{Li_Barrow_2010,Hellwing_etal_2010}, and showed a good agreement with the first results of \citet{Baldi_etal_2010}.

These could be summarized as follows:
\begin{itemize}
\item The DE-CDM interaction determines a faster growth of linear density perturbations as compared to $\Lambda $CDM, such that in order to
obtain the same normalization of the power spectrum amplitude at $z=0$ (\ie models with the same $\sigma _{8}$ at the present time) it is necessary to rescale 
the density fluctuations amplitude in the initial conditions with the appropriate growth factor for each different interacting DE model;
\item The selective interaction only between DE and CDM leaves the baryons completely uncoupled, and as a consequence baryonic and CDM
density fluctuations evolve with a different rate in interacting DE models; this determines a significant reduction of the halo baryon fraction of collapsed objects
at $z=0$;
\item If models are normalized at $z=0$ for what concerns the amplitude of their linear power spectrum, the mass function of halos has a comparable 
shape and amplitude at $z=0$ for all the models, while it differs at larger redshifts for different values of the coupling $\beta $; this does not hold if the models are normalized
at high redshift (as \eg with a normalization at CMB), in which case interacting DE models show a significant increase of massive objects at any
redshift as compared to $\Lambda $CDM, as shown by \citet{Baldi_Pettorino_2010};
\item For the case of constant couplings considered in this work, the CDM density profiles of massive halos at $z=0$ are always less concentrated in interacting DE scenarios
as compared to $\Lambda $CDM; this does not necessarily hold for the more general case of time dependent couplings \citep[see][]{Baldi_2010}.
\end{itemize}

Due to the complex interplay of all the new physical effects arising in interacting DE models and discussed in Sec.~\ref{cde} and \ref{mog},
it is an interesting exercise to try to isolate the individual impact that each of these different modifications of the standard gravitational
dynamics has on the final evolution of linear and nonlinear structures and to investigate how they separately contribute to each of the peculiar
features of interacting DE models listed above. 
This kind of analysis had already been carried out in \citet{Baldi_etal_2010} for the effect of concentration reduction in interacting DE models, by artificially switching off
in the N-body integration some of the new physical effects during the latest stages  of structure formation where nonlinear processes take place.
This has been done by computing the average formation redshift $z_{{\rm form}}$ of the most massive halos identified in the simulations and by 
selectively suppressing each specific effect of the DE-CDM interaction only after a redshift $z_{{\rm nl}}$ somewhat larger than $z_{{\rm form}}$. 
For instance, in \citet{Baldi_etal_2010} the average formation redshift in the sample of the 200 most massive halos formed in the different simulations
was found to be $z_{{\rm form}}\sim 0.8$, and the swhitch-off redshift for the test simulations was chosen to be $z_{{\rm nl}} = 1.5$.

This procedure ensures to isolate as much as possible the nonlinear effects of the interacting DE models from the linear evolution of the overall power spectrum
amplitude and therefore to be able to compare the properties of nonlinear objects that form in similar environments and with comparable formation histories.
The outcomes of the analysis carried out with this approach in \citet{Baldi_etal_2010} showed that 
the reduction of halo concentrations is primarily determined by the velocity-dependent acceleration (\ref{velocity_acc}) with a minor contribution
from the effect of mass variation of CDM particles. 

A more recent study carried out by \citet{Li_Barrow_2010b} has shown some partial
discrepancies with the early results of \citet{Baldi_etal_2010} claiming that the mass variation effect be actually the most relevant mechanism to 
determine the reduction of halo concentrations.
However, the analysis of \citet{Li_Barrow_2010b} did not adopt the above mentioned procedure aimed at distinguishing the linear and nonlinear 
impact of the different effects of the DE-CDM interaction, but rather switched off each individual effect right from the start of the numerical integrations.
This might determine some confusion in the interpretation of the results, and drive to misleading conclusions, as we will show below.

In the present work we significantly extend the analyses done by \citet{Baldi_etal_2010} and by \citet{Li_Barrow_2010b} and we try to address
the claimed discrepancies between the two treatments, clarifying the confusion that might arise when the linear and nonlinear effects
of interacting DE models are mixed with each other.

To this end, we will make use of the same code presented in \citet{Baldi_etal_2010}, to which we refer the interested reader for an 
extensive discussion of the implementation strategies. With this code we run a series of test simulations for an interacting DE model with $\beta = 0.24$ 
and with different numerical setups, and for a fiducial $\Lambda $CDM model\footnote{Please notice
the convention assumed for the definition of the coupling, which differs from what assumed in part of the literature. A coupling of $\beta =0.24$ with this
convention corresponds to $\beta =0.3$ in the notation of \citet{Amendola_2000}.}. 
Following the approach discussed in Sec.~\ref{norm_bckgd}, 
all the models are normalized to have the same set of cosmological parameters at $z=0$ in accordance to the most recent results from WMAP \citep[][]{wmap7}. 
A coupling as large as $\beta = 0.24$ is an extreme case and is already ruled out by several independent observational constraints \citep[see \eg][]{Bean_etal_2008,LaVacca_etal_2009,Xia_2009,Baldi_Viel_2010}, but our aim here is to amplify the impact of the interaction in order disentangle
more clearly the contributions arising from the different specific effects of interacting DE. This means that the simulations presented in this work are not intended to represent 
realistic models of the universe, but rather to provide a clear description of how interacting DE affects structure formation in the linear and in the nonlinear regimes.

Our simulations consist 
of a cosmological box of 80 comoving Mpc/h aside filled with $256^{3}$ particles for CDM and baryons for a total of $N=N_{c}+N_{b}\approx 3.4\times 10^{7}$ simulation particles.
This setup determines a spatial resolution of $\epsilon _{g}\approx 7 $ kpc/h, which is lower with respect to the work of \citet{Baldi_etal_2010} but comparable with \citet{Li_Barrow_2010b}. The baryons are included  to give a correct representation of the effective coupling, but 
in order to avoid spurious effects besides the ones determined by the DE-CDM interaction we do not include hydrodynamical forces in the simulations.
In other words, the baryonic particles are treated as an additional family of uncoupled collisionless particles which contribute to
reduce the total effective coupling with respect to the unrealistic case where the DE couples to all massive particles in the universe.
The mass resolution is $m_{c}(z=0)\approx 2.0\times 10^{9} $ M$_{\odot }/$h for the CDM particles and $m_{b} \approx 3.9\times 10^{8}$ M$_{\odot }/$h for the baryons.

With these numerical parameters we run a series of 8 simulations in which we artificially switch off one of the characteristic effects of interacting DE,
adopting both the procedures used by \citet{Baldi_etal_2010} and by \citet{Li_Barrow_2010} in order to compare the outcomes of these test runs
with the full interacting DE model and with the fiducial $\Lambda $CDM cosmology. 
It is of particular relevance for our discussion to notice here that the former approach ensures to only slightly perturb the
normalization of the background cosmological parameters and of the linear perturbations amplitude of the different 
test runs with respect to the fiducial $\Lambda $CDM scenario at $z=0$. On the contrary, the latter procedure significantly
changes the normalization of the background cosmological parameters at $z=0$, and more importantly shifts
by a large offset the linear amplitude of density perturbations in the different runs, such that each of the test simulations will have a
significantly different value of $\sigma _{8}$ at $z=0$, thereby making quite difficult a direct comparison of the nonlinear
properties of the collapsed objects forming in the different runs.
The details of these simulations are described in Table~\ref{tab:simulations}.
\begin{center}
\begin{table*}
\label{tab:simulations}
\begin{center}
\begin{tabular}{ccccccl}
\hline
Name & Model & Modified Hubble function & fifth-force & velocity-dependent acceleration & mass variation & mass normalization\\
\hline
\hline
S0 & $\Lambda $CDM & {\bf inactive} & {\bf inactive} & {\bf inactive} & {\bf inactive} & $m_{c}(z) = m_{c}(0) \forall z$ \\
S1 & $\beta = 0.24$ & active & active & active & active & $m_{c}(z) > m_{c}(0) \forall z > 0$ \\
S2 & " & {\bf inactive} & active & active & active & $m_{c}(z) > m_{c}(0) \forall z > 0$ \\
S3 & " & active & {\bf inactive} & active & active & $m_{c}(z) > m_{c}(0) \forall z > 0$ \\
S4 & " & active & active & {\bf always inactive}  & active & $m_{c}(z) > m_{c}(0) \forall z > 0$ \\
S5 & " & active & active & active & {\bf always inactive} & $m_{c}(z) = m_{c}(\infty ) \forall z$ \\
S6 & " & active & active & active & {\bf always inactive} & $m_{c}(z) = m_{c}(0) \forall z$ \\
S7 & " & active & {\bf inactive for $z < 2$} & active & active & $m_{c}(z) > m_{c}(0) \forall z > 0$\\
S8 & " & active & active & {\bf inactive for $z < 2$} & active &  $m_{c}(z) > m_{c}(0) \forall z > 0$\\
S9 & " & active & active & active & {\bf inactive for $z < 2$} & $\left\{ \begin{minipage}{100pt}
$m_{c}(z) > m_{c}(2) \forall z > 2$\\
$m_{c}(z) = m_{c}(2) \forall z <= 2$
\end{minipage} \right. $\\
\hline
\end{tabular}
\caption{Table of the different simulations presented in this work. S0 corresponds to the standard $\Lambda $CDM model, S1 to the full interacting DE simulation. Simulations S2-S6 adopt the same procedure used by \citet{Li_Barrow_2010b} while simulations S7-S9 follow the method of \citet{Baldi_etal_2010}.}

\end{center}
\end{table*}
\end{center}

\begin{figure}
\begin{center}
\includegraphics[scale=0.4]{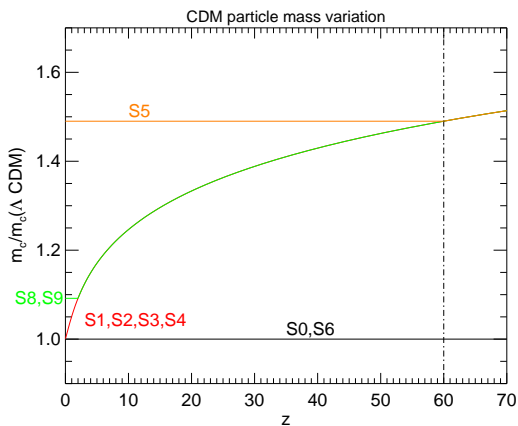}
  \caption{(Color online) The evolution of CDM particle mass as a function of redshift in all the simulations of Table~\ref{tab:simulations}. The curved line correponds to the evolution in the full interacting DE model. This evolution can be stopped at different times in the different simulations according to the different numerical methods discussed in the text.}
\label{fig:mass_variation}
\end{center}
\end{figure}

Simulations S2-S5 correspond to the procedure adopted by \citet{Li_Barrow_2010b}, to which we add simulation S6 where the mass variation
of CDM particles is also suppressed during the whole run but the mass is normalized to the value it has at $z=0$ in the other simulations. 
These two different normalizations of the CDM particle mass
will determine a huge difference in the final results, as we will show below. 
In fact, simply switching off the mass variation of CDM particles from the starting redshift of the simulations
but leaving the particle mass normalization to the same initial value of the full simulations would result in a much larger value of $\Omega _{c}$ at $z=0$ as
compared to the fiducial $\Lambda $CDM model, while this would not happen if the mass is normalized to the value it takes at $z=0$ in the other runs.
It therefore comes as no surprise that in the former case the global amplitude of linear density perturbations will grow significantly faster than in the full interacting DE model
represented by the simulation S1, and 
significantly slower in the latter one. The situation is well explained in Fig.~\ref{fig:mass_variation}, where the mass evolution of CDM particles is shown 
for all the simulations of Table~\ref{tab:simulations}. 
\normalsize
Clearly, adopting a mass normalization to the value that the CDM mass has at $z=60$ in the full model (as done in the simulation S5 and in the analysis by \citet{Li_Barrow_2010b}) forces a significantly larger 
$\Omega _{c}$ during the whole expansion history of the universe, and as a consequence a much faster linear growth of perturbations that will influence the subsequent nonlinear evolution
of massive halos. For instance, with this approach halos will form much earlier and in a universe with a much larger CDM density than they would do in the full simulation. 
It is therefore again not surprising
that halos forming in this kind of simulation, as found by \citet{Li_Barrow_2010b}, will have a larger characteristic overdensity and a somewhat larger halo concentration.
However, this is just an artifact arising from the different effective cosmological parameters that are forced into the simulation by normalizing the mass at high redshifts, and not by the direct effect
of the mass variation onto nonlinear structures.
In other words, this effect will be primarily a consequence of the different mass normalizations rather than of the mass constancy in time.

The same argument applies to the case of simulation S6, where the mass is normalized to the value of $m_{c}(z=0)$ in the full simulation. In this case structures will evolve with a
lower effective $\Omega _{c}$ than in the full simulation, that will determine a slower growth of density perturbations. As a consequence, starting from a lower amplitude with respect to $\Lambda$CDM in the initial conditions, the CDM density perturbations of simulation S6 will not have the time to catch up the same $\sigma _{8}$ of the other models at $z=0$, thereby determining a later formation time of massive halos.

A similar situation concerns the suppression of the velocity-dependent acceleration. As we explained above, this term has completely different effects in
the linear and nonlinear regimes of structure formation. In the former situation, in fact, this term always accelerates the growth of linear density perturbations
thereby acting in the same direction of a larger value of $\Omega _{c}$, while in the latter it can significantly reduce nonlinear overdensities at small scales.
This means that suppressing this term right from the beginning of the simulations has a similar effect as what discussed above for the mass normalization, and 
could lead to a different amplitude of linear perturbations at $z=0$ as compared to the fiducial $\Lambda $CDM and the full interacting DE model.

On the contrary, suppressing the mass variation and the velocity-dependent acceleration only at recent epochs (in this analysis we have chosen $z_{nl}=2$), where the nonlinear stages of structure formation take place (as done for simulations S8 and S9) ensures
that the CDM density and the amplitude of linear density perturbations at $z=0$ be comparable with both the $\Lambda $CDM reference model and the full interacting DE simulation. This was in fact the procedure adopted by the first work of \citet{Baldi_etal_2010}, and allows a direct comparison of the impact that each different effect of the DE-CDM interaction has on the nonlinear regime of structure formation only, without spurious effects induced by different background or linear evolutions of the models.

It is nevertheless important to stress here that this type of study is only a way to better understand how the DE-CDM interaction acts on the linear and nonlinear stages of structure formation,
while it does not represent any realistic situation. The only realistic scenario is the full model where all the above mentioned
effects are simultaneously and consistently included. For instance, the mass variation and the velocity-dependent acceleration
are two aspects of the same phenomenon, namely the conservation of particles' momentum, and none of the two can be in place without the other.
It is nevertheless interesting to perform this kind of analysis to get a deeper insight into the dynamics of coupled systems.

In the next Section we will present the outcomes of the whole set of simulations of Table~\ref{tab:simulations} both concerning the linear and nonlinear
stages of structure formation, and we will show that the assessed partial disagreement between the results of \citet{Baldi_etal_2010} and \citet{Li_Barrow_2010b} 
is actually just a consequence of the procedure adopted in the latter study for the suppression of the individual effects under investigation,
which inevitably mixes linear and nonlinear effects of the DE-CDM interaction.

\section{Results}
\label{res}

We now compare the results of our test simulations with the reference $\Lambda $CDM model and with the full implementation
of interacting DE carried out in simulations S0 and S1, respectively.
We consider several different observable quantities related to the linear and the nonlinear evolution of structures
and we try to identify which of the various effects of the DE-CDM interaction have a more prominent 
impact on each of these observables.

\subsection{Matter Power Spectrum}

\begin{figure*}
\begin{center}
\includegraphics[scale=0.32]{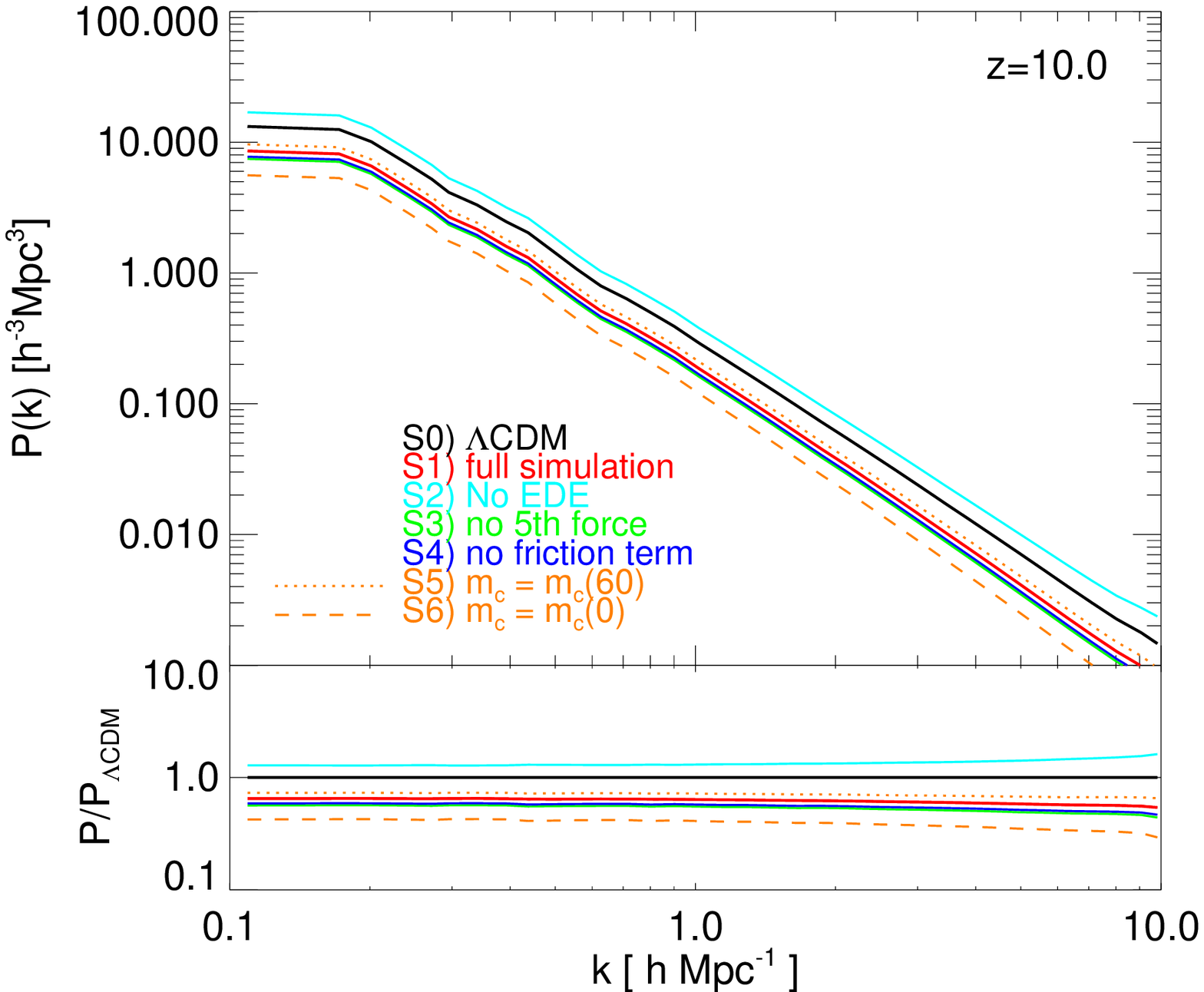}
\includegraphics[scale=0.32]{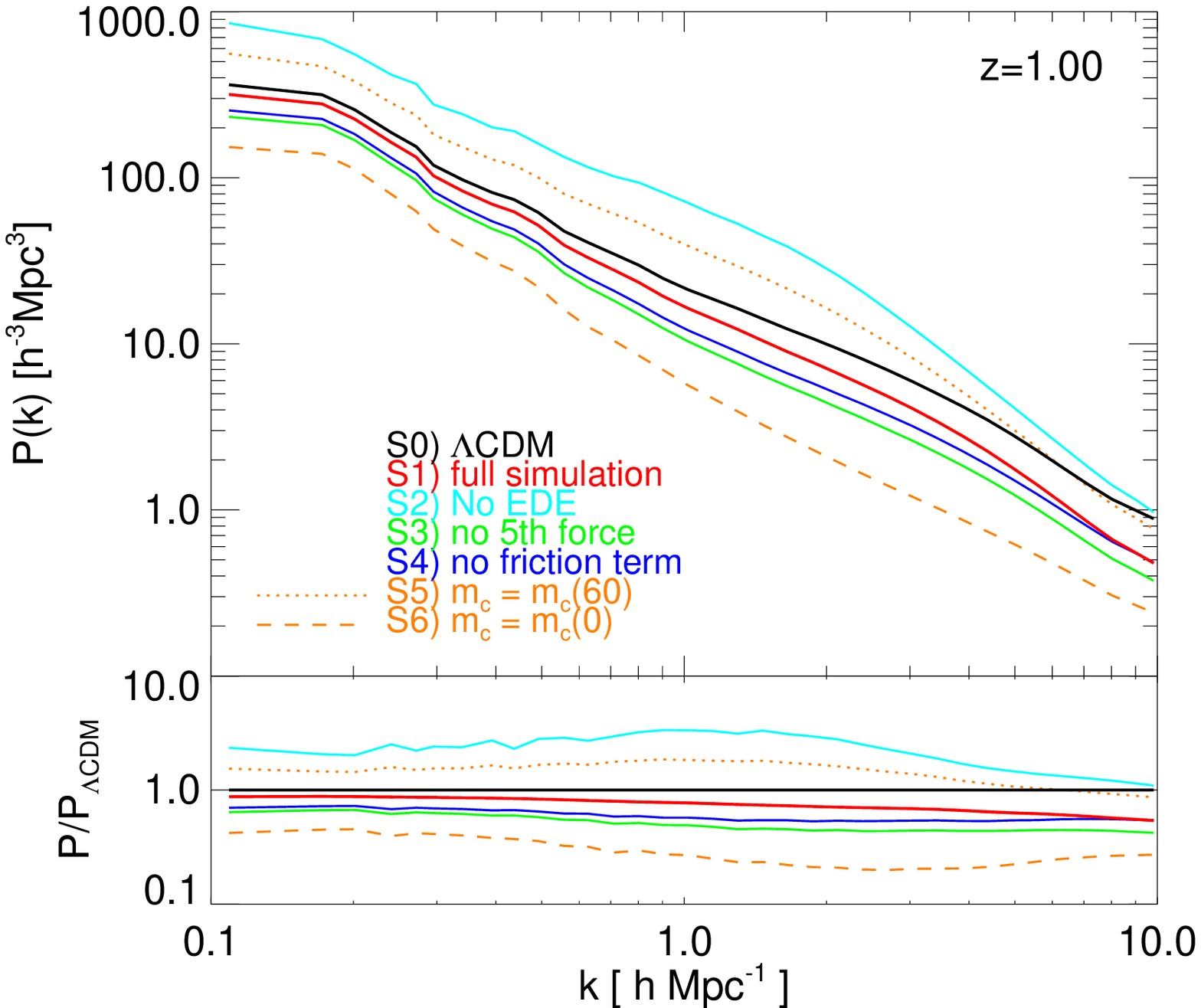}
\includegraphics[scale=0.32]{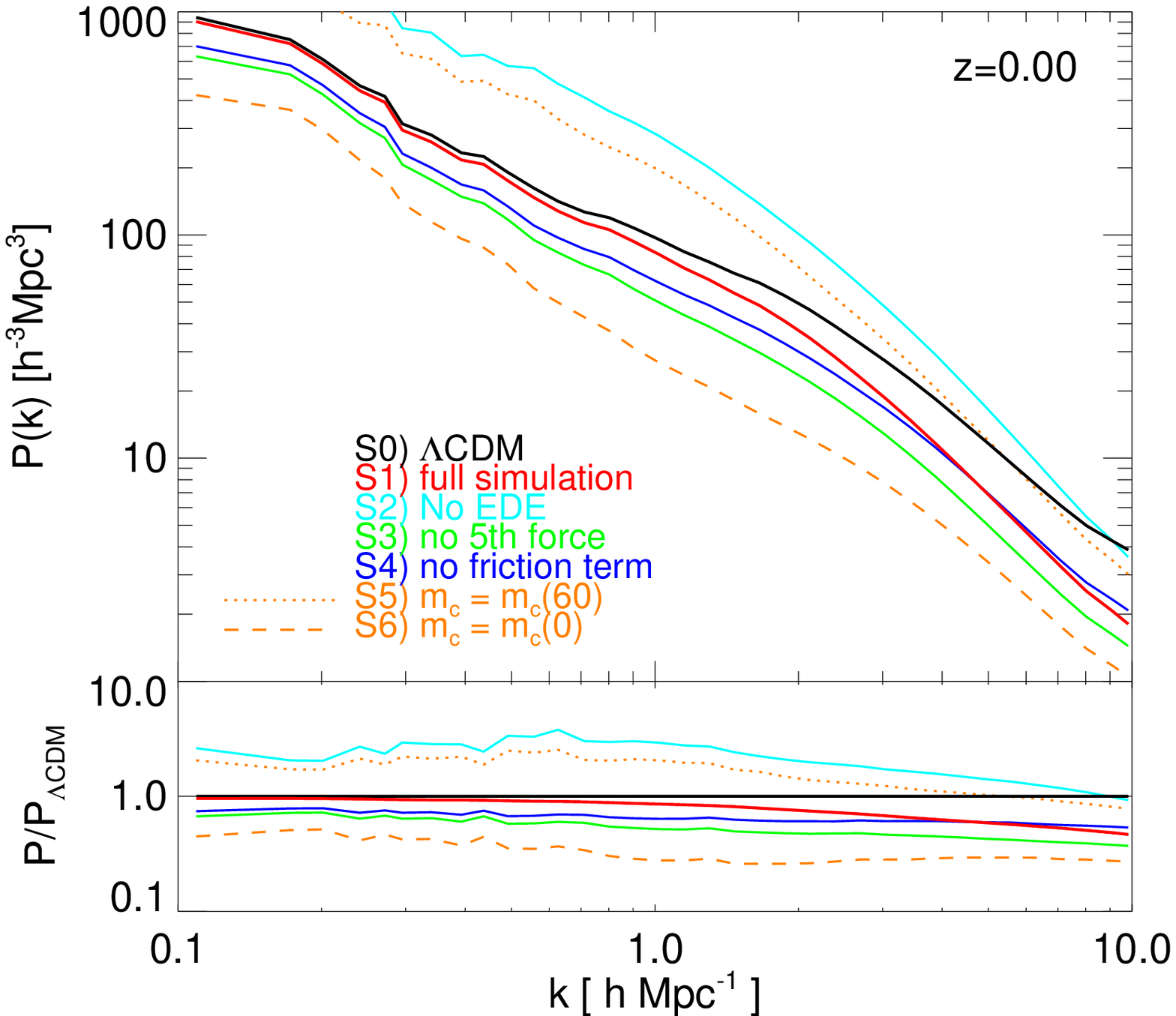}\\ 
\includegraphics[scale=0.32]{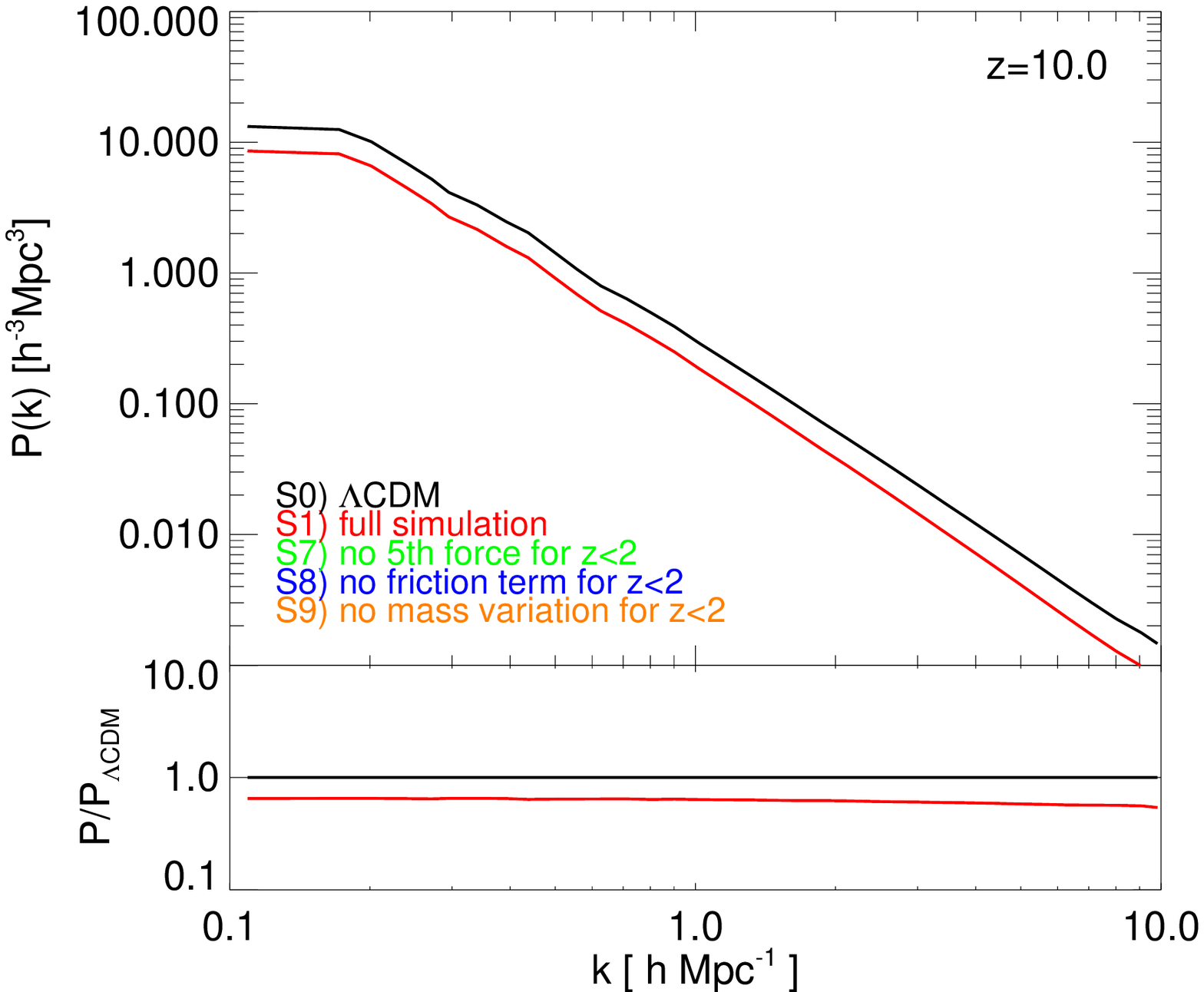}
\includegraphics[scale=0.32]{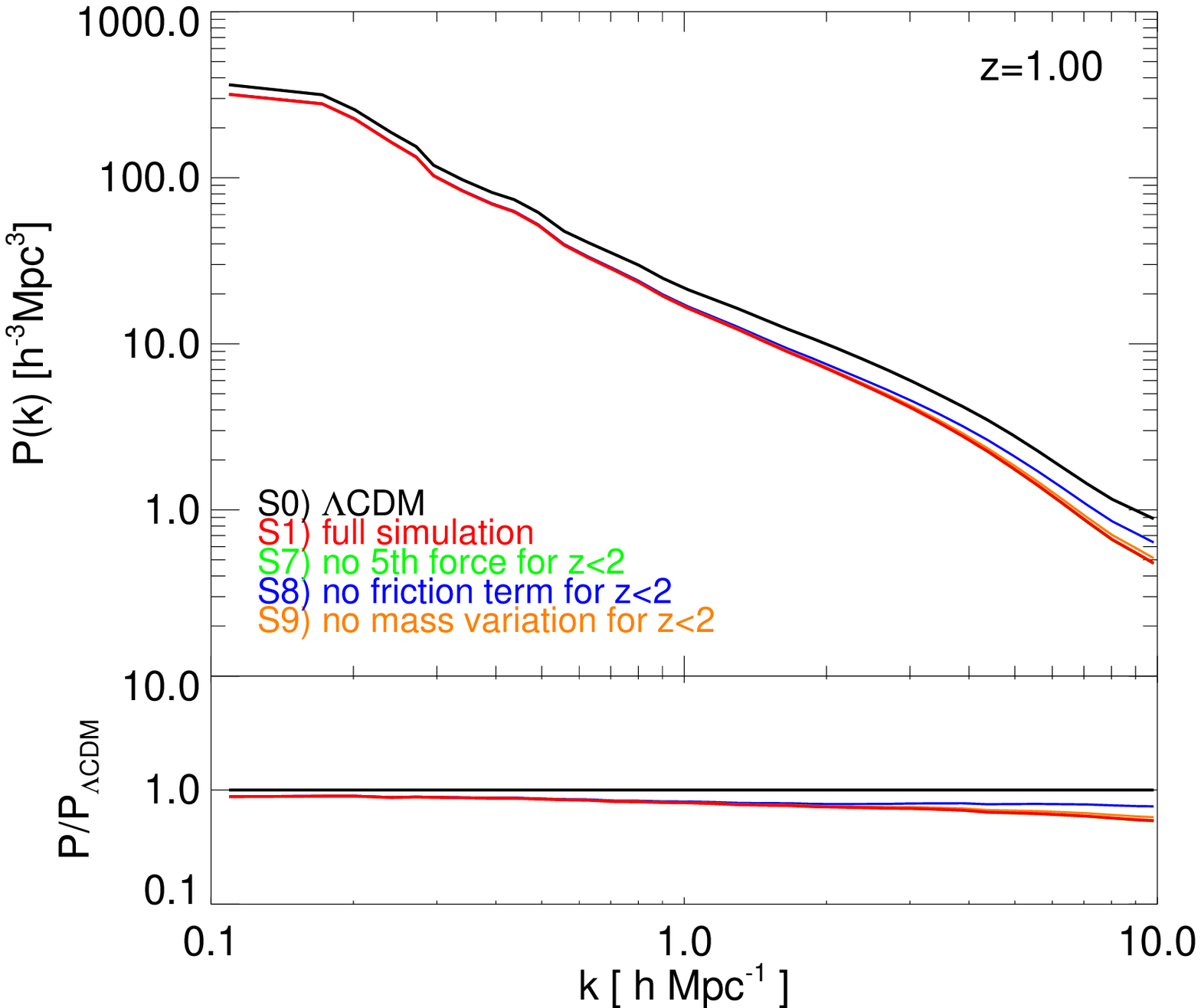}
\includegraphics[scale=0.32]{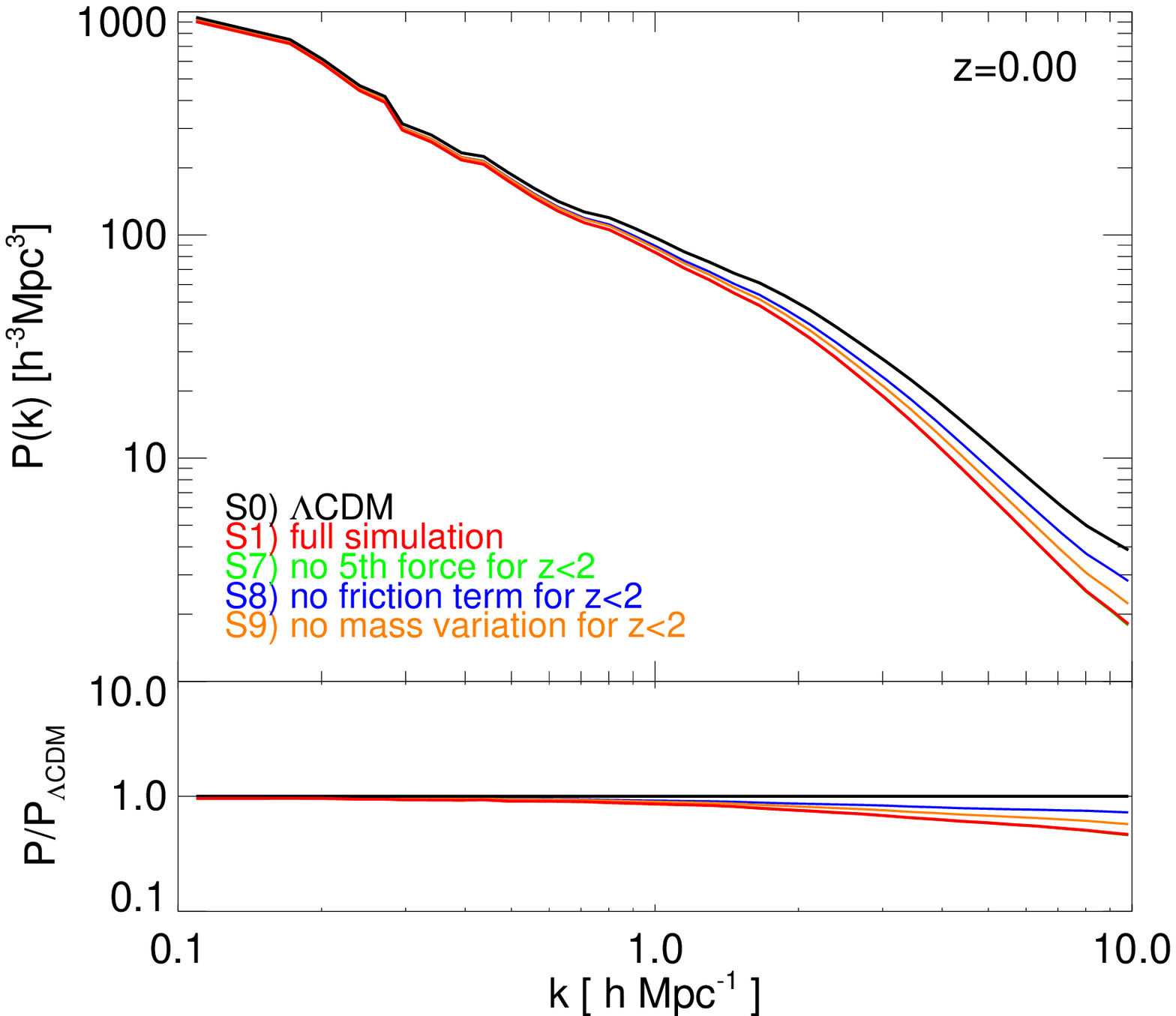} 
  \caption{(Color online) The matter power spectrum of $\Lambda $CDM (black) and interacting DE (red) as a function of inverse scale $k$ at different redshifts. The upper panels show the test simulations run suppressing specific effects of the DE-CDM interaction right from the start of the simulations. Lower panels show the results of simulations with a suppression only during the latest stages of structure formation, \ie for $z\le z_{{\rm nl}}=2$. Clearly, the large scatter present in the upper panels witnesses the superposition of linear and nonlinear effects arising as a consequence of the procedure adopted in simulations S2-S6. The method used in simulations S7-S9 is clearly more suitable to compare the nonlinear effects of interacting DE models, and shows how the velocity-dependent acceleration (blue) is the most relevant mechanism in the suppression of small-scale power in these models.}
\label{fig:power_spectra}
\end{center}
\end{figure*}
\normalsize

As a first step, we compute for all of our simulations the total matter pawer spectrum $P(k,z)$ as a function of inverse scale $k$ at different redshifts $z$,
and the separate power spectra for CDM and baryonic particles, $P_{c}(k,z)$ and $P_{b}(k,z)$, respectively.
In Fig.~\ref{fig:power_spectra} we show the total matter power spectrum $P(k,z)$ at redshifts $z=10,1,0$. The upper panels refer to the simulations
where the different individual effects of the DE-CDM interaction have been suppressed from the start of the simulation at $z_{i}=60$ \citep[as done in the analysis of][]{Li_Barrow_2010b}, while the lower panels refer to the procedure of switching off the specific effects only at late times, \ie at $z \leq z_{{\rm nl}} = 2$, as for the
study of \citet{Baldi_etal_2010}.

First of all, one can clearly identify how the DE-CDM interaction affects the total matter power spectrum by comparing the black ($\Lambda $CDM) and the red (interacting DE)
solid curves in the $z=0$ plots. While the two power spectra have (by construction) the same amplitude at the largest scales, \ie are normalized to give the same linear perturbations
amplitude at $z=0$, at smaller and smaller scales (\ie larger values of $k$), the interacting DE model features a progressively stronger lack of power
as compared to $\Lambda $CDM. This already shows how the linear and the nonlinear regimes are very differently affected by the DE-CDM interaction, since nonlinear scales
show a significant reduction of power while linear scales remain practically unaffected.

We can now compare how the different effects of the DE-CDM interaction individually contribute to the total distortion of the matter power spectrum by having a look at how suppressing
each specific effect influences the power spectrum at different $z$. Already at a first glance, it is easy to realize how the top panels of Fig.~\ref{fig:power_spectra} present
a much larger scatter of the different curves at all scales, including the largest scales available in the simulation box, as compared to the bottom panels.
This already clearly shows how suppressing any of the characteristic effects of the DE-CDM interaction from the starting of the simulations (as done in the upper panels) determines 
a large change in the respective values of $\sigma _{8}$ at the present time. In particular, suppressing the background modification of the expansion history 
(\ie replacing the proper expansion history of the interacting DE model with the standard $\Lambda $CDM one) produces a faster growth
of density perturbations that consequently reach a higher normalization at the present time, as shown by the cyan line in the top panels of 
Fig.~\ref{fig:power_spectra}. This is fully consistent with the fact that 
the interaction determines a larger value of the Hubble function at $z > 0$ (as shown in the right panel of Fig.~\ref{fig:background}), and 
therefore replacing it with the $\Lambda $CDM
Hubble function determines a slower expansion rate as compared to the full simulation. This result is however exactly the opposite of what shown by
\citet{Li_Barrow_2010b}, where a different normalization has been used and where interacting DE models feature a slower expansion rate than $\Lambda $CDM at $z > 0$.

The suppression of mass variation, as we discussed above, also determines a different effecive $\Omega _{c}(z)$ throughout the simulation, thereby 
inducing a faster or slower growth of density perturbations depending on whether the constant mass value is fixed to the initial ($m_{c}=m_{c}(z_{i})$, simulation S5, dotted orange line) or to the final ($m_{c}=m_{c}(0)$, simulation S6, dashed orange line) value taken in the full simulation S1, respectively. This effect clearly appears
in the upper panels of Fig.~\ref{fig:power_spectra}, where the high- and the low-redshift normalizations of the CDM particle mass give rise to higher and lower 
values, respectively, of the power spectrum amplitude at all scales.

The suppression of the fifth-force (green lines) and of the velocity-dependent acceleration (blue lines) produce weaker effects on the overall amplitude of the
power spectra, while it is already clear from the upper panels of Fig.\ref{fig:power_spectra} that the velocity-dependent acceleration has a stronger impact at nonlinear scales
since its suppression (simulation S4) clearly shows an increase of power at large values of $k$ with respect to the full simulation S1 which is not detected if the
fifth-force is suppressed (simulation S3).

The analysis of the nonlinear effects becomes then much more clear if we move to analyze the 
outcomes of test simulations run adopting the procedure of \citet{Baldi_etal_2010}. As we stressed above, this procedure seems more suitable if one wants to compare
the nonlinear effects of interacting DE models since it does not alter the overall linear normalization of the different test simulations under study.
This is very clear by looking at the lower panels of Fig.~\ref{fig:power_spectra}. At $z > z_{{\rm nl}}$ all the simulations show the same power spectrum at all scales, as 
expected, with a lower overall amplitude as compared to $\Lambda $CDM due to the scaling of the initial conditions. At $z \le z_{{\rm nl}}$ the different test simulations start to deviate from each other at small scales, while the large scale normalization stays the same until $z=0$, where it finally coincides also with the $\Lambda $CDM one.

This allows to compare the importance of the different effects of the DE-CDM interaction on the nonlinear scales for structures embedded in a density field
with the same linear amplitude normalization at large scales. This more meaningful comparison already shows a clear hierarchy of the relative contribution of the
different effects to the total distortion of the nonlinear matter power spectrum. If the fifth-force (green) seems to have basically no effect in this context, the 
velocity-dependent acceleration (blue) is clearly the most important effect in erasing power at small scales, while the mass variation (orange) has a sizeable but subleading
impact. This result is in contrast with what shown by \citet{Li_Barrow_2010b}, and shows how a high-redshift normalization can determine a potentially
misleading superposition of linear and nonlinear effects.

\subsection{Baryon bias}

\begin{figure*}
\begin{center}
\includegraphics[scale=0.32]{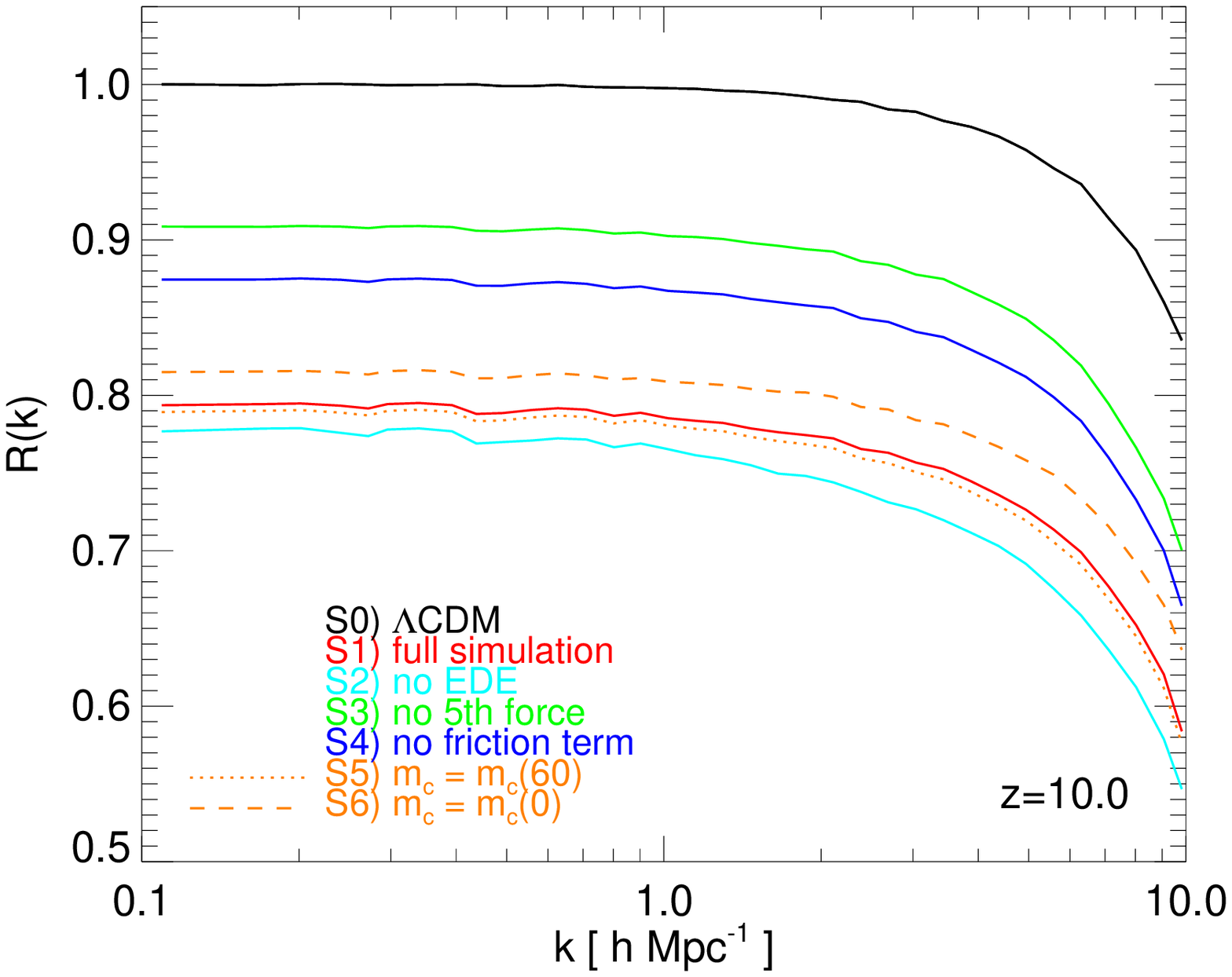}
\includegraphics[scale=0.32]{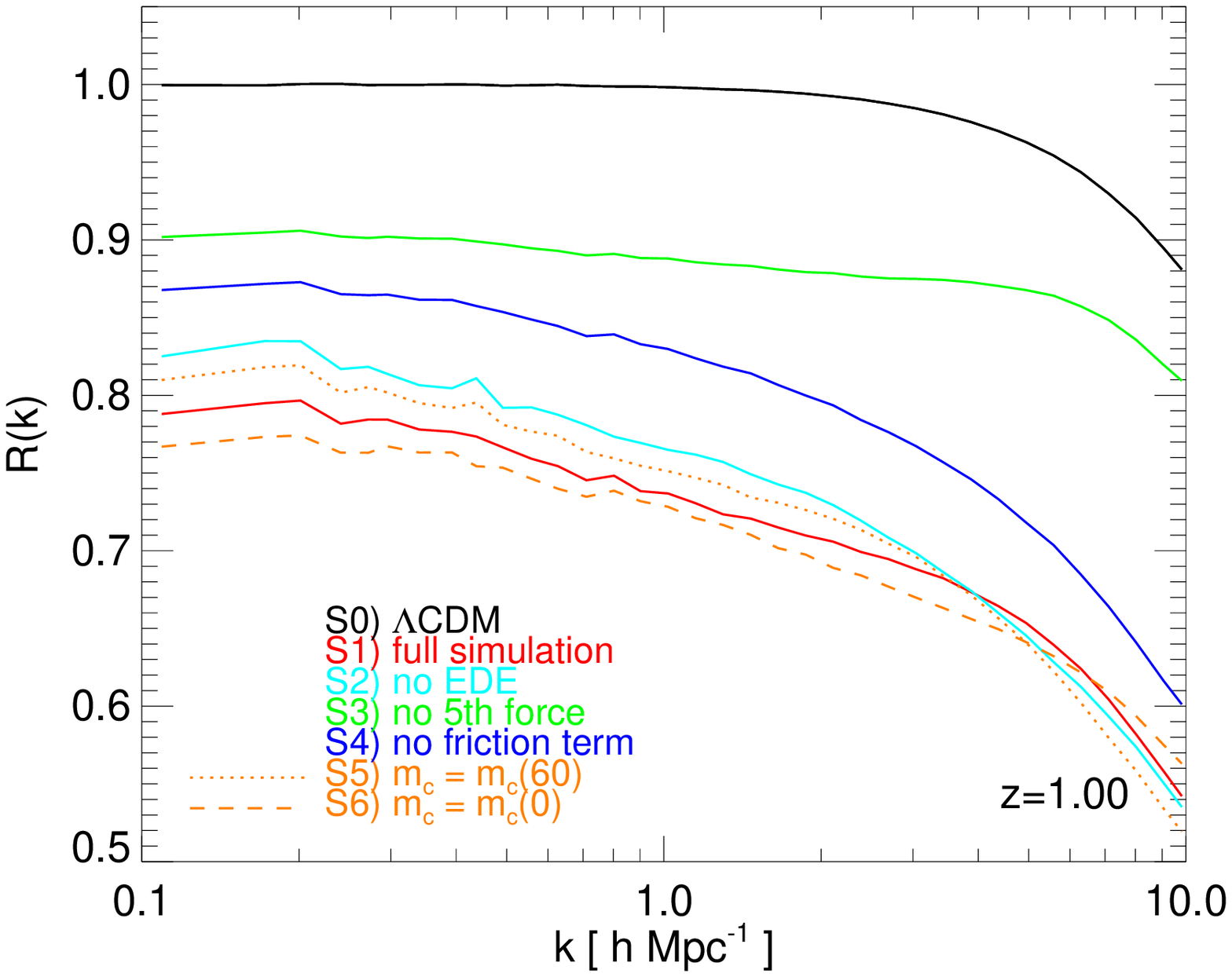}
\includegraphics[scale=0.32]{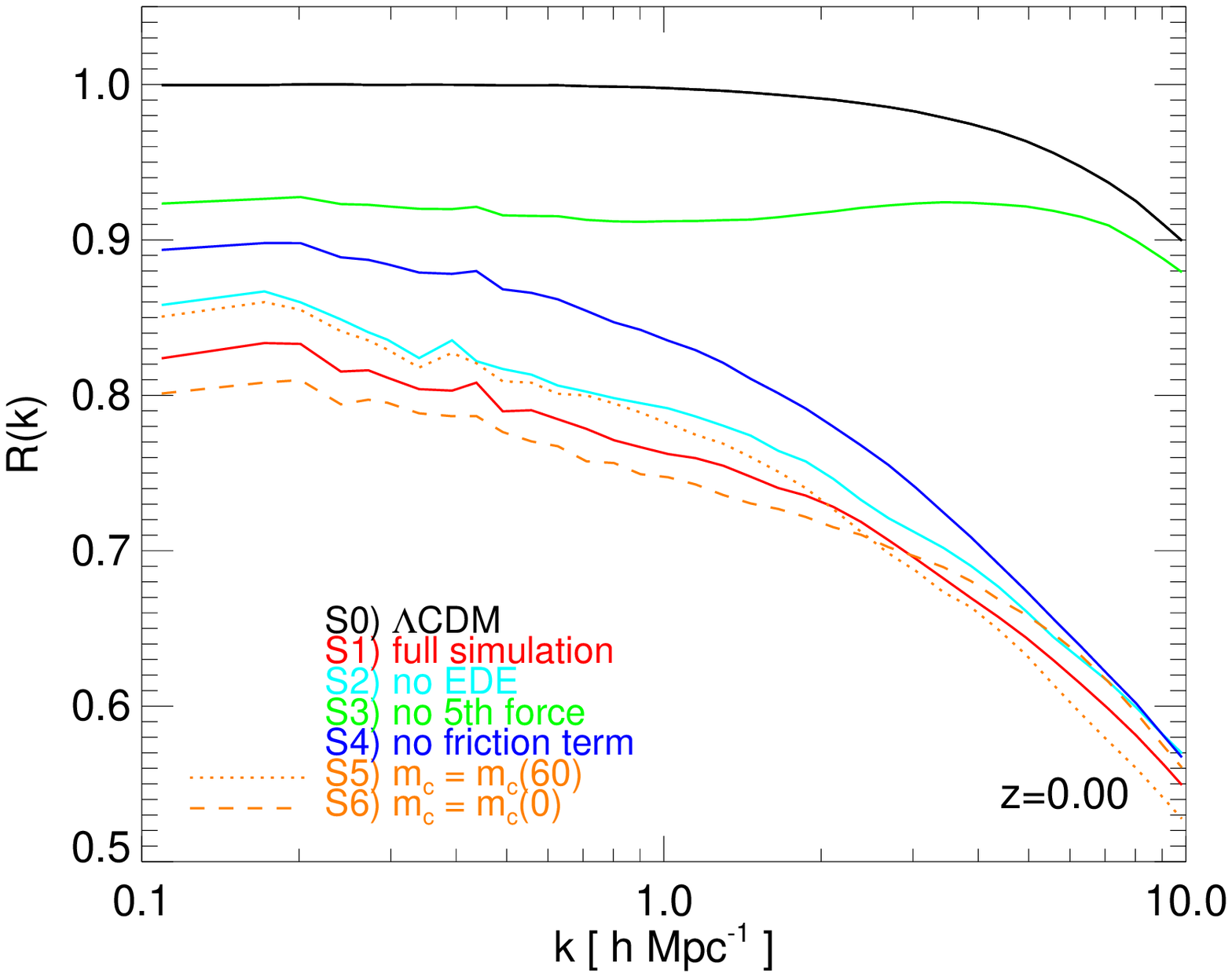}\\ 
\includegraphics[scale=0.32]{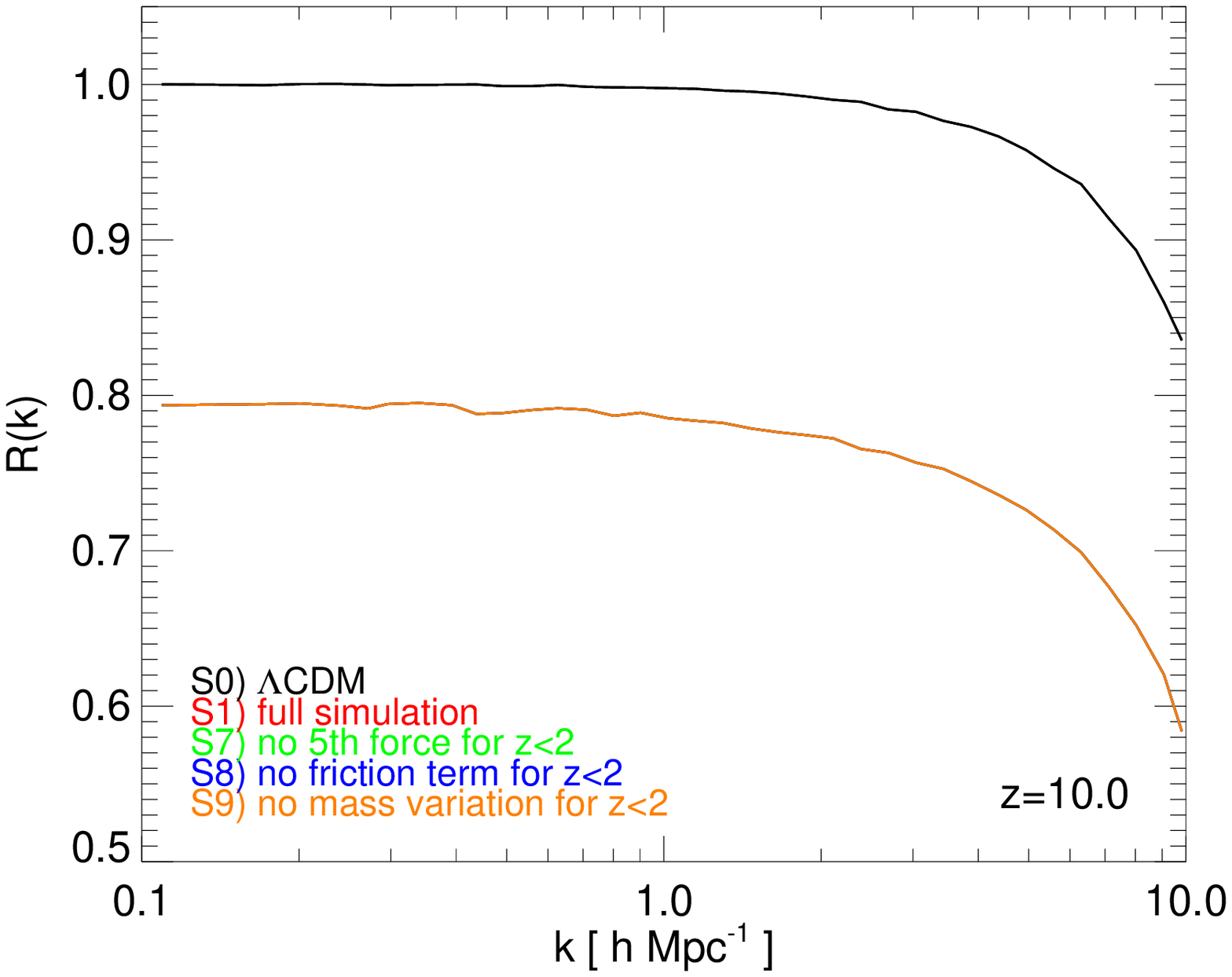}
\includegraphics[scale=0.32]{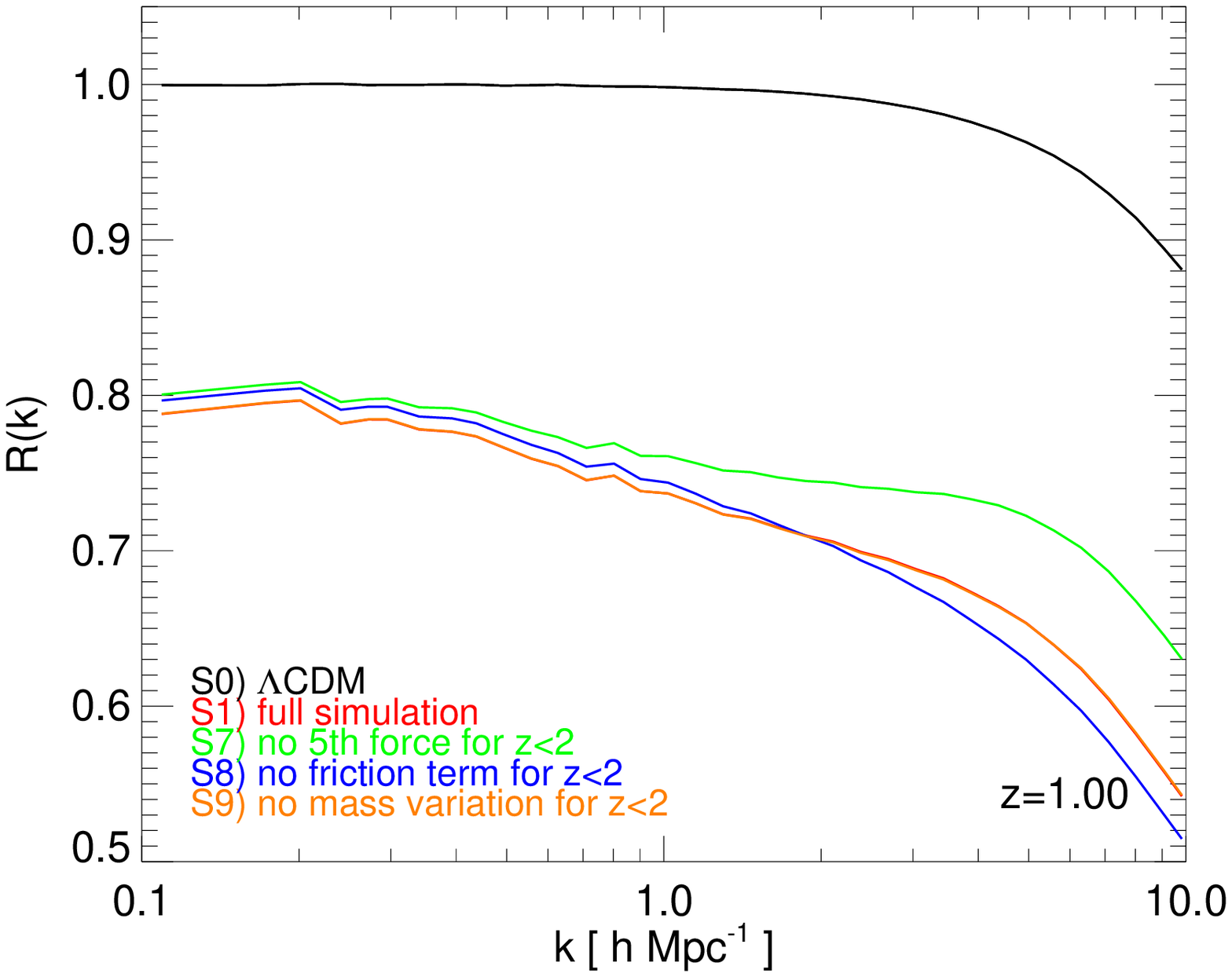}
\includegraphics[scale=0.32]{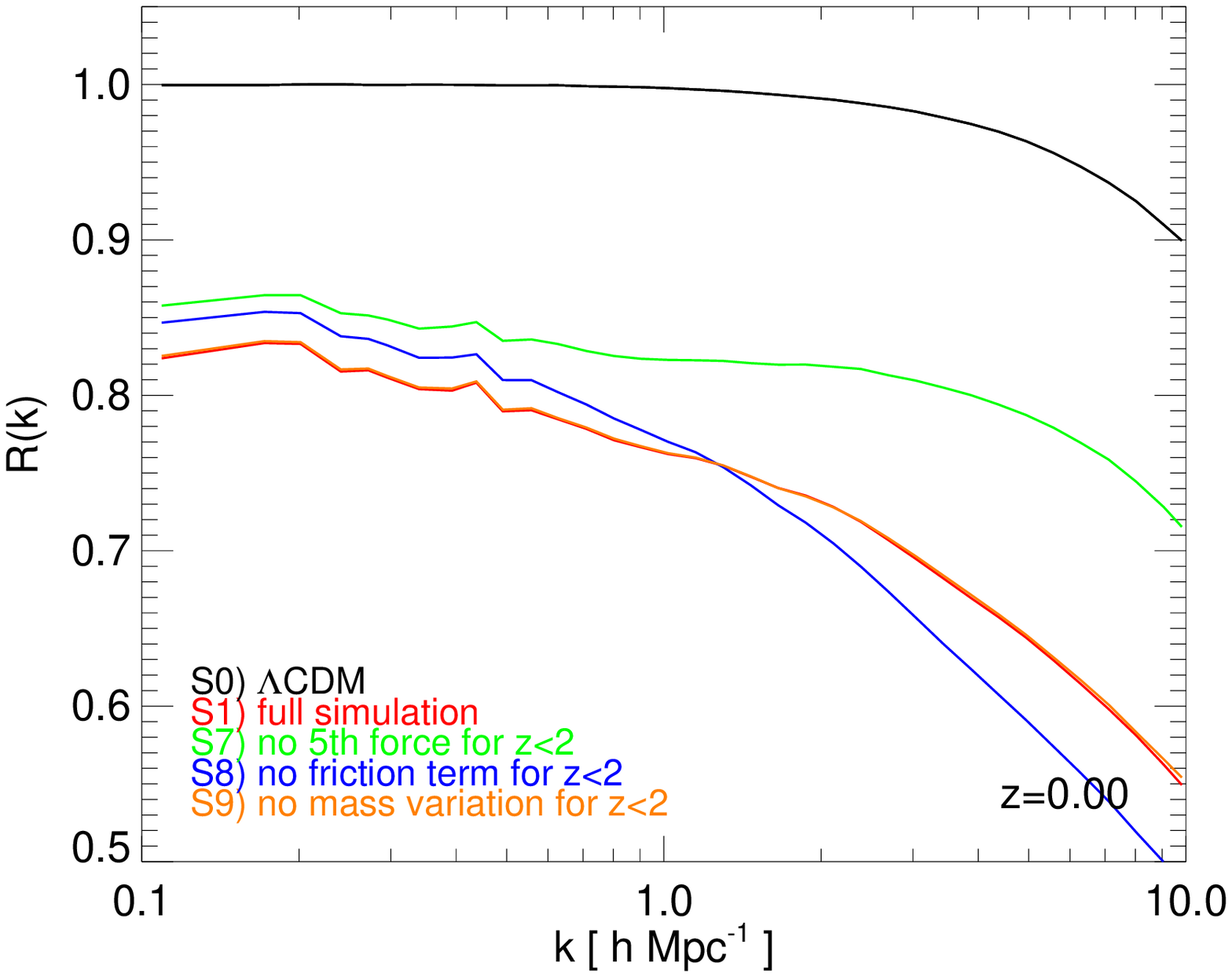}
  \caption{(Color online) The evolution of the gravitational bias $R(k,z)$ as a function of inverse scale $k$ for the $\Lambda $CDM model (black) and the interacting DE scenario (red). Upper plots show the results of the test simulations S2-S6 with high-redshift suppression of the individual effects, while lower panels show the results of simulations S7-S9 with suppression only for $z\le z_{{\rm nl}}=2$. The leading role of the fifth force (green) in determining a gravitational bias clearly appears in both sets of simulations.}
\label{fig:bias}
\end{center}
\end{figure*}
\normalsize

We now analyze the impact of the different individual effects of the DE-CDM interaction on a quantity that
has been called in the literature the ``gravitational bias" \citep[][]{Baldi_etal_2010}, represented by the ratio
of the baryon to CDM power spectra defined as $R(k,z)\equiv P_{b}(k,z)/P_{c}(k,z)$.
This quantity shows how the density perturbations of baryons and CDM, which start with the same relative
amplitude in the early universe, evolve during structure formation processes. Clearly, in absence of hydrodynamic
forces acting on baryons (as it is the case for the numerical works under investigation) the only difference in the growth of density perturbations between baryons and CDM
is given by the DE-CDM interaction, such that for $\Lambda $CDM one expects $R(k,z)=1\, \forall k,z$. This is indeed the case if one looks at Fig.~\ref{fig:bias}, where the
ratio $R(k,z)$ is plotted at different redshifts for all the simulations under study. Again, upper panels are for the simulations where the individual effects have been switched 
off right from the start, while the lower panels are for the case of switch-off only for $z \le z_{{\rm nl}}$.

Also in this case the upper panels show a much larger scatter of the different simulations as compared to the lower panels, due to the large linear effects
arising from an early suppression of the interacting DE effects. However, it is particularly interesting to notice in both types of approaches how the fifth-force
term, that showed very little effects on the total power spectrum shown in Fig.~\ref{fig:power_spectra} and discussed in the previous section, is now the
leading mechanism in determining the evolution of the gravitational bias both at linear and nonlinear scales,
such that its suppression significantly shifts up the value of $R(k,z)$ towards the $\Lambda $CDM value of $1.0$.
Since the evolution of the gravitational bias is tightly related to the baryon fraction of massive collapsed objects as galaxy clusters \citep[as shown in ][]{Baldi_etal_2010},
our analysis shows how the fifth-force term, far from being an irrelevant effect for interacting DE models, could actually determine observational features in the
baryonic budget of groups and clusters of galaxies.

It is then also very interesting to notice the impact of the velocity-dependent acceleration. Although the upper panels of Fig.~\ref{fig:bias}, due to the large scatter induced by the high redshift suppression of the interacting DE effects, do not allow a clear detection of this behavior, the lower panels distinctively show how the velocity-dependent acceleration
differently affects the linear and the nonlinear regimes of structure formation, as discussed in Sec.~\ref{mog}. In fact, while in the linear regime probed by the largest scales
in the simulation the effect of the velocity-dependent acceleration is to enhance the growth of CDM density perturbations, in the small scale regime it has the opposite 
effect of reducing nonlinear power. As a consequence, the suppression of this acceleration (that acts only on CDM particles) has the 
effect of increasing the value of the gravitational bias at large scales towards the $\Lambda $CDM value as compared to the full interacting DE model, 
and of further reducing it at small scales, as clearly 
represented by the crossover of the red line (full simulation) and the blue line (simulation S8) at $k\sim 1$ h Mpc$^{-1}$ in the last panel of Fig.~\ref{fig:bias}.
This behavior distinctly shows that the velocity-dependent acceleration cannot be treated in the same way in the linear and nonlinear regimes of structure formation,
and that a suppression of this term already at high redshifts will inevitably determine a superposition of its linear and nonlinear impact onto the final
results of the simulations, making it very difficult to disentangle the two different contributions.

\subsection{Halo Mass Function}

\begin{figure*}
\begin{center}
\includegraphics[scale=0.48]{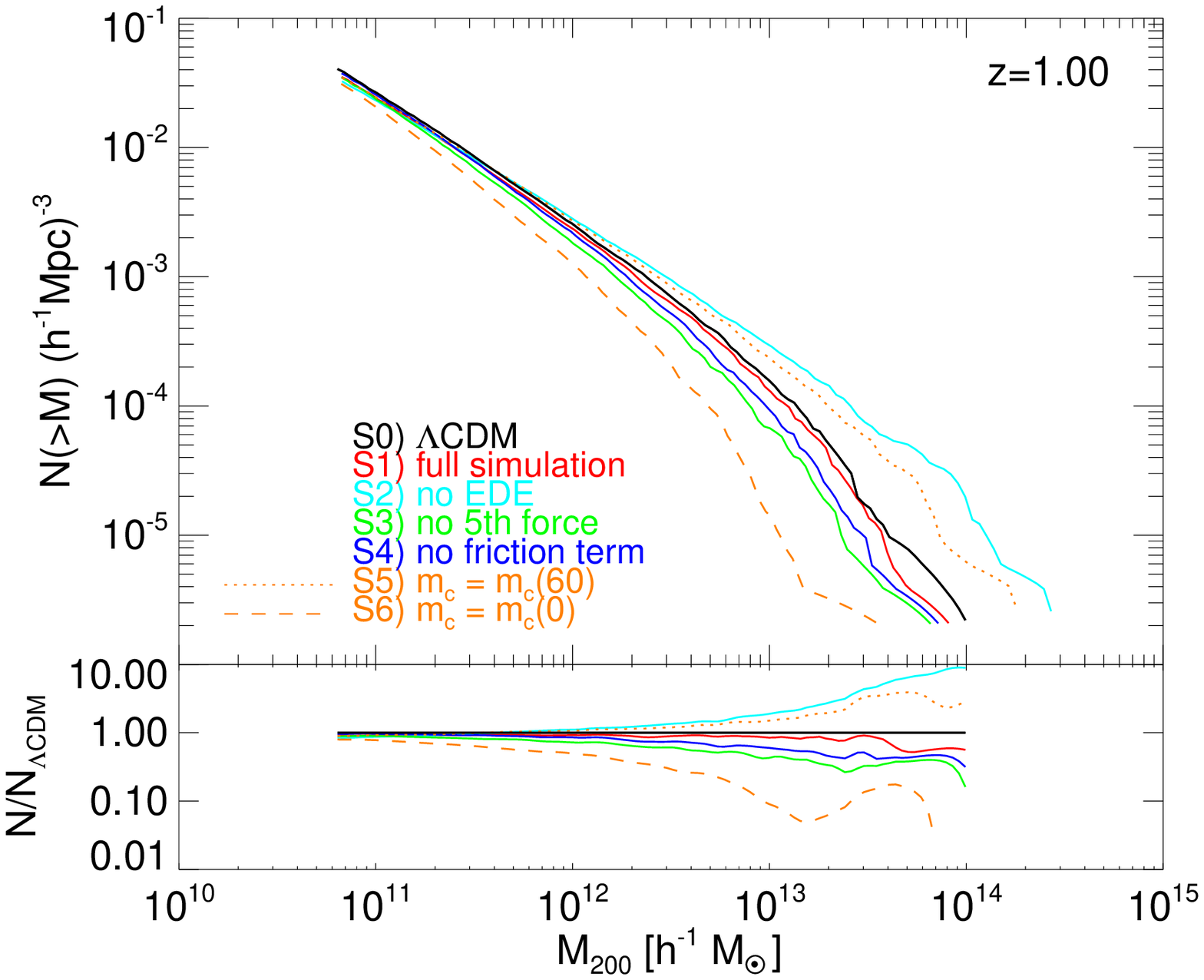}
\includegraphics[scale=0.48]{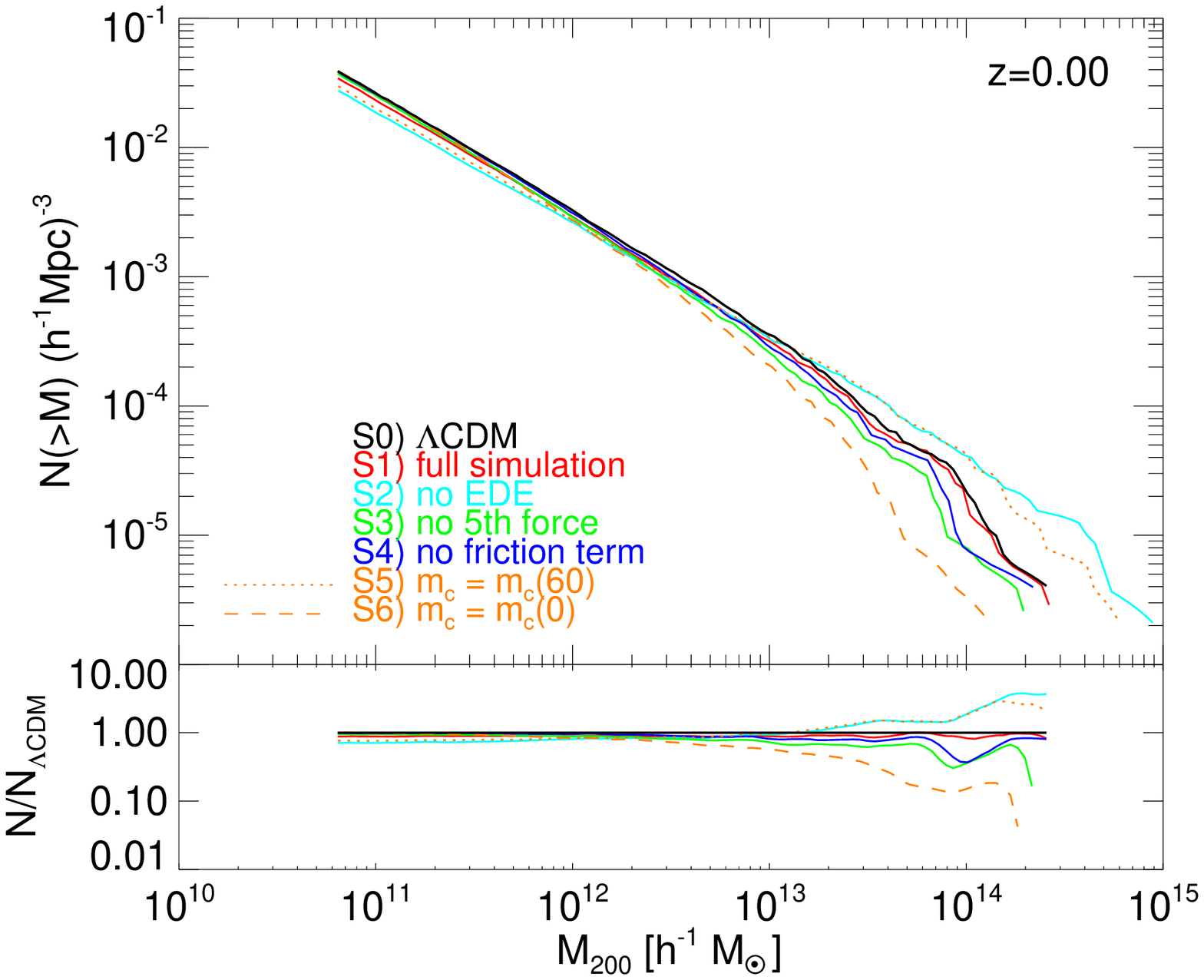}\\
\includegraphics[scale=0.48]{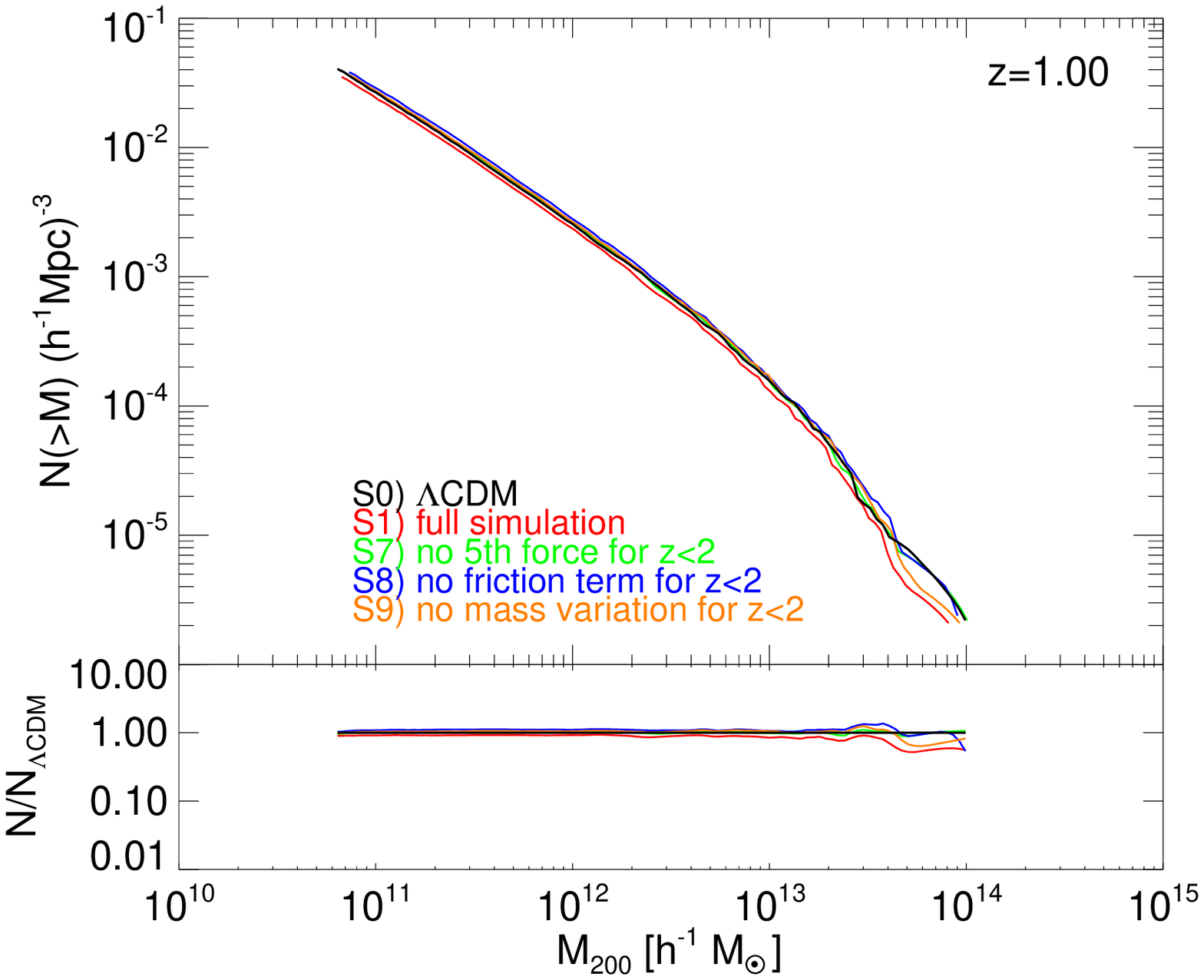}
\includegraphics[scale=0.48]{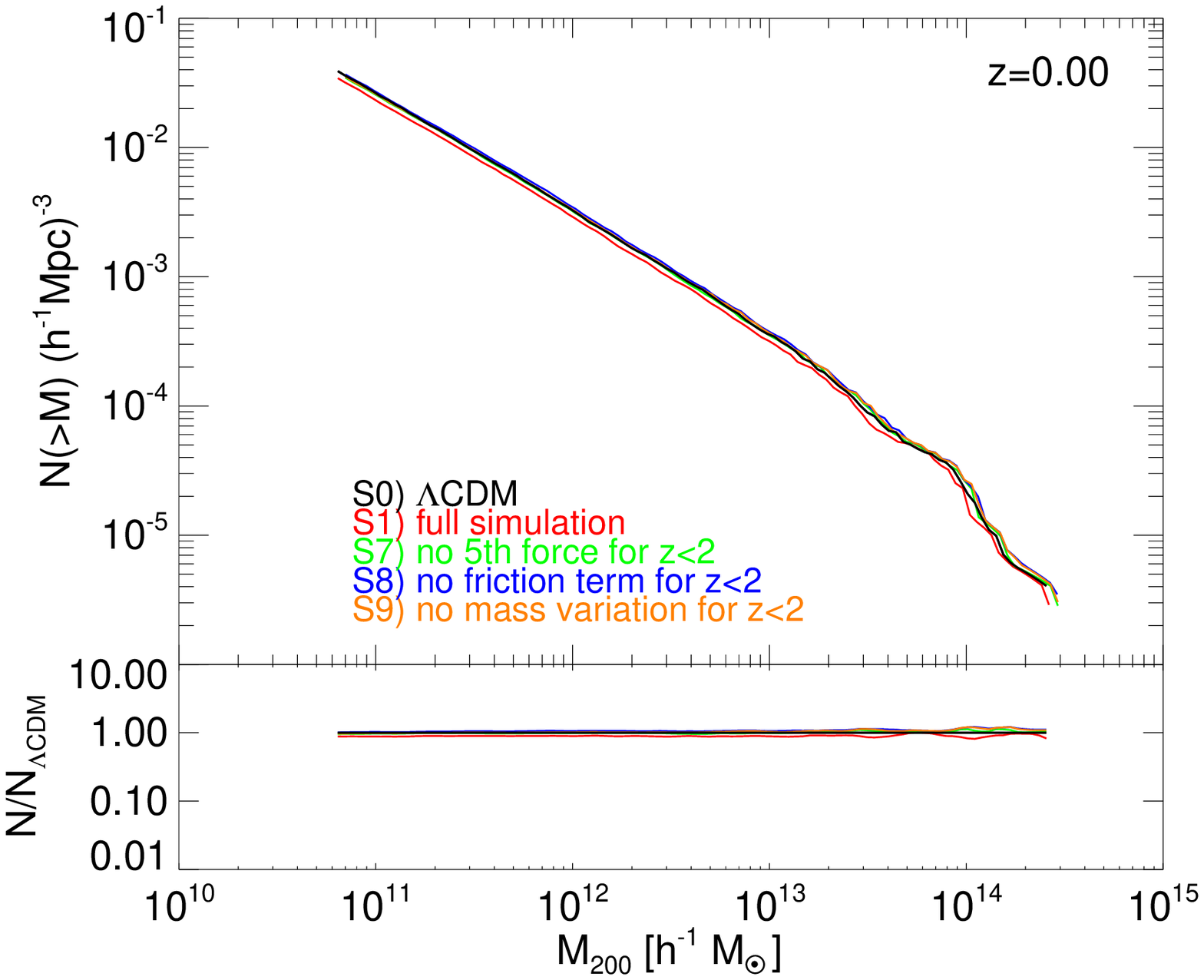}
  \caption{(Color online) The cumulative mass functions of CDM halos in $\Lambda $CDM (black) and interacting DE (red) models at $z=1$ and $z=0$.
  The large scatter of the different halo mass functions in the upper panels is due to the high-redshift suppression of the individual effects of the DE-CDM interaction in the simulations S2-S6. The same scatter does not appear if a low-redshift suppression is adopted.}
\label{fig:massfunction}
\end{center}
\end{figure*}
\normalsize

We now compare the impact that each of the effects under study has on the halo mass function at $z=0$ and at $z=1$.
To do so, we first identify the halos formed within each of the simulations described in Table~\ref{tab:simulations} by means of a 
Friends-of-Friends algorithm with linking length $\lambda = 0.2 \times \bar{d}$, where $\bar{d}$ is the average particle spacing.
We then identify halo substructures by means of the {\small SUBFIND} algorithm \citep{Springel2001}, and we compute
the cumulative CDM halo mass function as a function of the halo virial mass $M_{200}$.
The result is shown in Fig.~\ref{fig:massfunction} for the test simulations with a high-redshift switch-off of the different interacting DE effects
(upper panels) and for the test simulations with $z_{{\rm nl}} = 2$ (lower panels).

Also in this case, as expected, suppressing the effects of the DE-CDM interaction at high redshift determines a large scatter of the different test simulations at 
recent epochs, due to the significantly different linear growth history of the various runs. On the contrary, a low-redshift suppression of the individual effects
produces basically no scatter in the halo mass functions, which roughly follow the behavior of the full interacting DE model (red line) that by construction
is normalized in order to give a comparable distribution of structures to the corresponding $\Lambda $CDM model (black line) at $z=0$.

The comparison of these two different scenarios is particularly relevant for the considerations that we will make in the next Section. In fact, it clearly appears from
the upper panels of Fig.~\ref{fig:massfunction} that all the simulations with high-redshift suppression of the interacting DE effects will present very different structures
at the present time, with significantly different masses and with possible large offsets in the final location of the individual halos, and this deviation
from simulation to simulation is clearly more prominent at large masses. This will make particularly difficult to directly compare the properties of individual halos
within this set of simulations, since it will be more difficult to identify two halos formed in different simulations as the same object, as we will discuss in detail in the next Section.

This again shows the potentially misleading superposition of linear and nonlinear effects that characterizes the approach
of suppressing the new physical effects of interacting DE already at high redshifts in the simulations. The most suitable approach for studying  the nonlinear regime of interacting DE
models is therefore the one first adopted by \citet{Baldi_etal_2010} that allows to clearly disentangle the impact that each of the interacting DE effects has on the inner dynamics of nonlinear objects from the overall linear growth history. In fact, as we will show in the next Section, although the simulations displayed in the lower panels of Fig.~\ref{fig:massfunction} have roughly  the same statistic distribution of massive halos, the internal properties of individual halos still get significantly affected by each of the specific effects under investigation.

\subsection{Halo density profiles}

\begin{figure*}
\begin{center}
\includegraphics[scale=0.48]{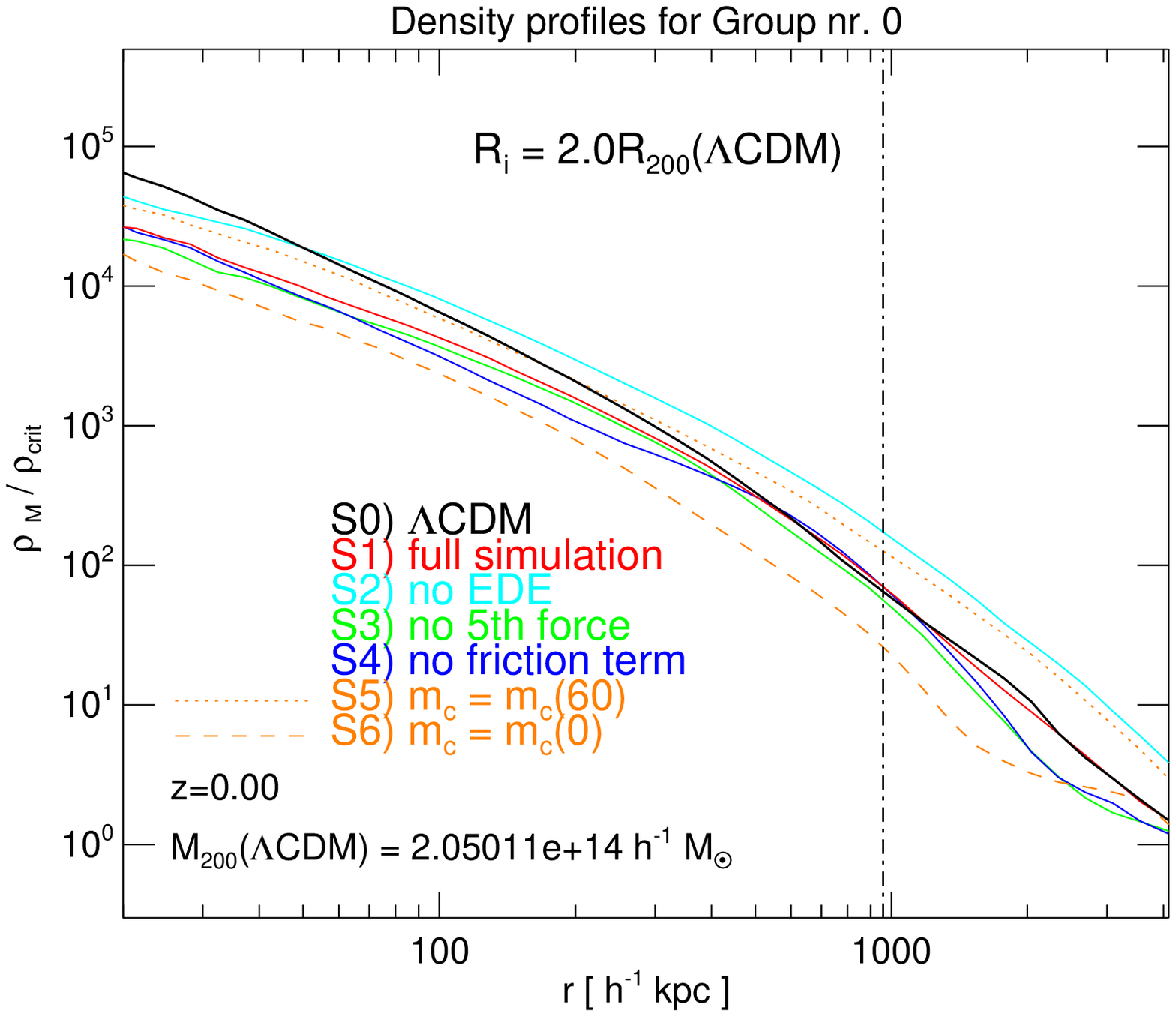}
\includegraphics[scale=0.48]{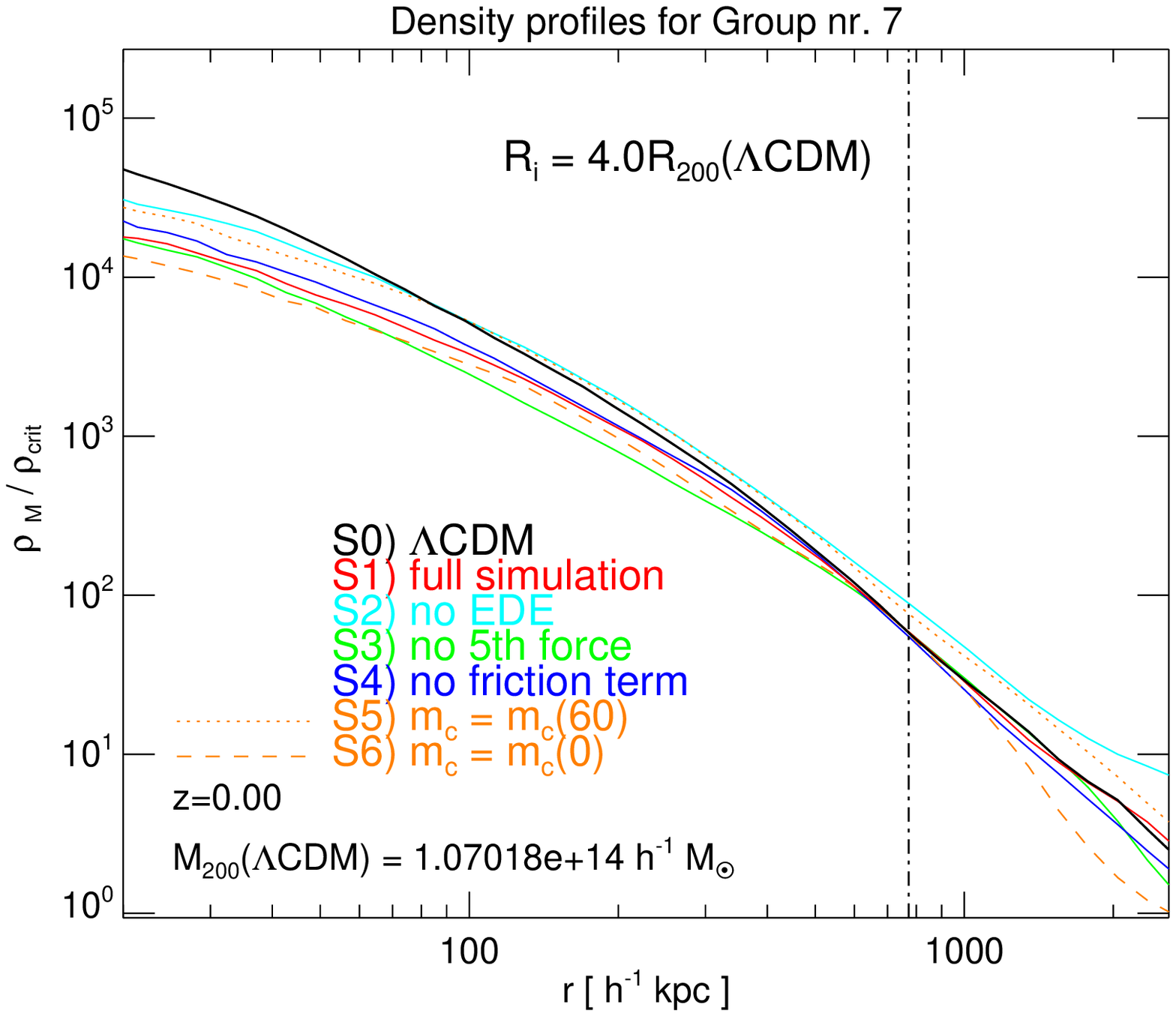}\\
\includegraphics[scale=0.48]{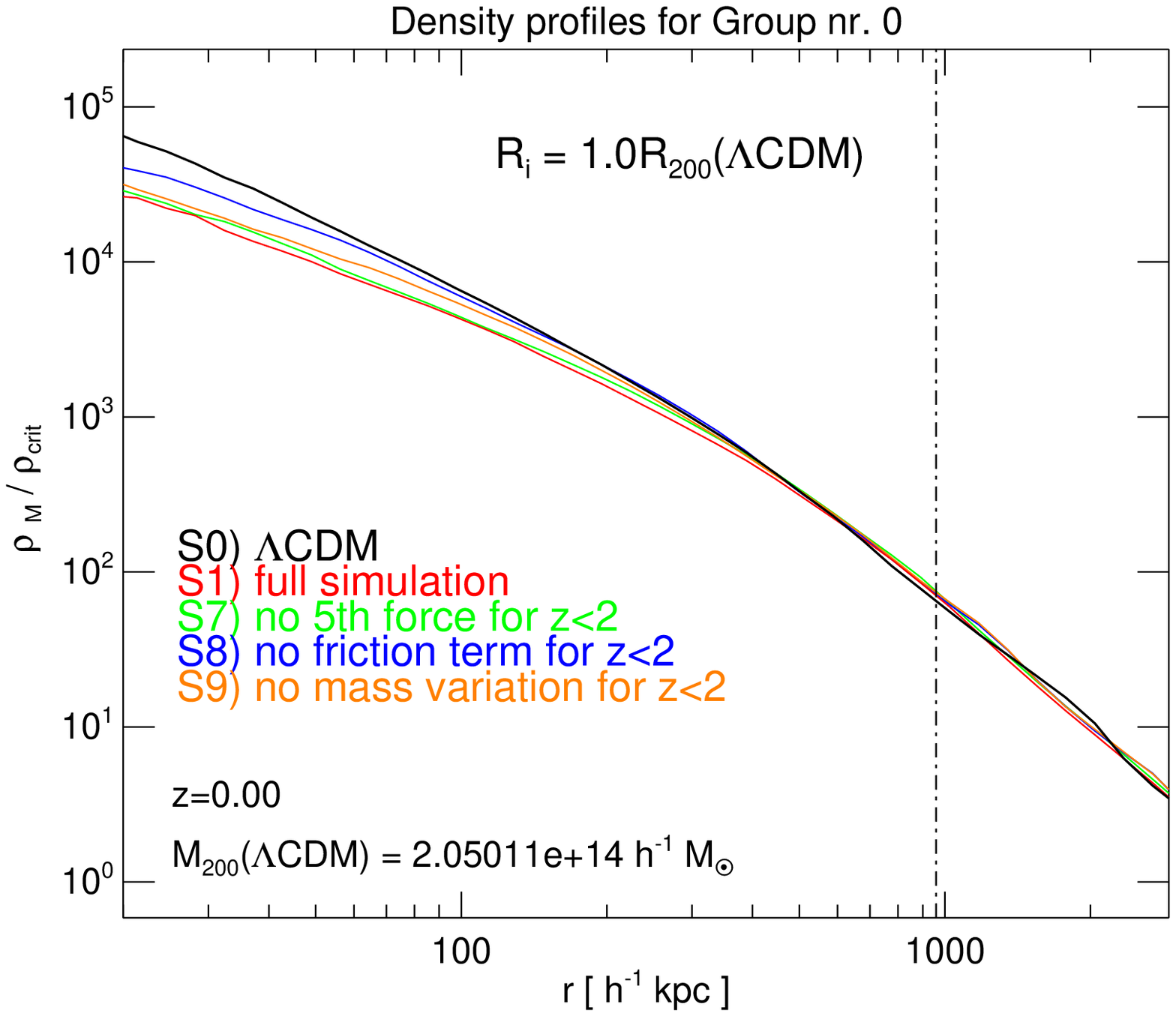}
\includegraphics[scale=0.48]{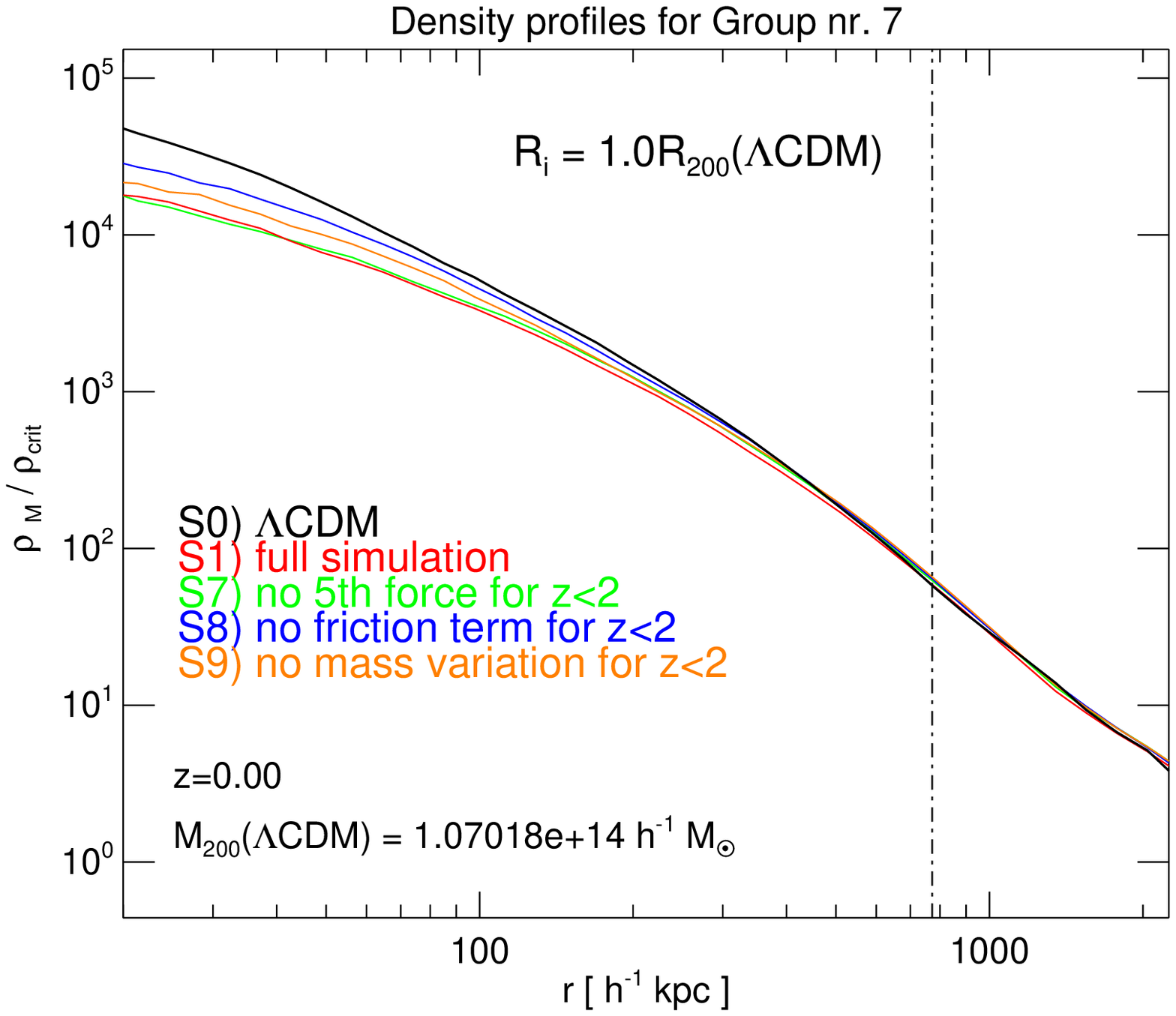} 
  \caption{(Color online) Matter density profiles of two massive halos in our sample for the $\Lambda $CDM (black) and interacting DE (red) models. As in the other figures, upper panels refer to the procedure which adopts a high-redshift suppression of the individual effects of the DE-CDM interaction, while lower panels refer to the suppression only for $z\le z_{{\rm nl}}=2$. The large scatter present in the former case does not allow a clear comparison of the relative importance of each individual effect on the nonlinear density profiles due to a superposition of linear and nonlinear effects. On the contrary, the lower panels clearly show that the velocity-dependent acceleration (blue) is the most important effect in driving the reduction of the inner overdensity of halos, in full agreement with the results of \citet{Baldi_etal_2010}.}
\label{fig:profiles}
\end{center}
\end{figure*}
\normalsize

The last and probably most interesting test of our interacting DE models concerns the internal dynamics of highly nonlinear structures as
the massive CDM halos that form in each of the different numerical realizations under study.
Since all the simulations are started from the same random realization of the linear power spectrum in the initial conditions, it is expected that
structures will form in the same position in all the simulations and this should allow a direct comparison of individual halos that formed in each of the runs.
However, this correspondence cannot be exact due to the different physical effects included in each simulation, and some offset in the final location
of the halos is expected as a consequence of the specific dynamical evolution of the systems in each run. 

It is nevertheless very important to stress here that if models are normalized in order to have the same linear perturbations amplitude at the present time,
the offset in the position of the halos will generally be a fraction of their virial radius $R_{200}$, and the local environment, the formation and merging history, and the dynamical
state of two corresponding halos in different simulations will be in general very similar. This has been tested by \citet{Baldi_etal_2010}, where it has been shown that
roughly 40\% of the 200 most massive halos in a set of four hydrodynamical N-body simulations with different values of the DE-CDM coupling $\beta $ showed an offset
smaller than one virial radius with respect to the location of their corresponding $\Lambda $CDM halo.

On the contrary, if different models have a significantly different normalization of the linear perturbations amplitude at $z=0$, the final structures formed in each simulation
will significantly differ from run to run, and will very likely be considerably offset from each other, as we showed in the previous section by comparing
the mass functions of halos at $z=0$ and $z=1$ in the simulations with a high-redshift suppression of the different interacting DE effects (upper panels of Fig.~\ref{fig:massfunction}).

Here we will apply a similar identification criterion to the one adopted in \citet{Baldi_etal_2010}, and within a given set of simulations we will identify with each other only those halos for which the location of
the most bound particle (taken as the center of the halo) lies within a radius $R_{i}=\eta R_{200}$ from the center of a corresponding $\Lambda $CDM structure.
We can directly confirm our expectations by counting which fraction of halos can be identified with a given $\Lambda $CDM object by using different values of $\eta $
for the identification radius $R_{i}$, first within the set of simulations
that do not provide a common normalization of the linear perturbations amplitude at $z=0$ (simulations S2-S6) and then within the simulations that as a result
of a low redshift suppression of the individual interacting DE effects determine a substantially identical normalization at the present time (simulations S7-S9).

For the former models, in fact, none of the halos at $z=0$ can be identified as being the same object in all the simulations when $\eta $ is set to unity.
This fraction increases to 1\% for $\eta = 2 $ and to 7\% and 17\% for $\eta =3 $ and $\eta = 4$, respectively. However, one should notice that 
an offset of $4R_{200}$ corresponds
for the most massive halos in our sample to a distance of $1.5-4.0$ Mpc/h between the halos centers, which clearly does not guarantee that these halos can be
safely considered as the same object forming in the different simulations.

On the other hand, the latter models, which adopt the same procedure of \citet{Baldi_etal_2010}, already
have a fraction of 80\% of identifiable halos for $\eta =0.5$, which increases to 91\% for $\eta = 1$.
It is therefore clear that these models will provide a much more homogeneous and statistically significant sample of halos for 
a direct investigation of the purely nonlinear effects of interacting DE models.

This is confirmed by having a look at Fig.~\ref{fig:profiles}, where the density profiles of two halos in our sample are plotted 
for the S2-S6 simulations with high redshift suppression of the specific interaction effects (in the upper panels) and for the 
S7-S9 simulations with suppression only for $z\le z_{{\rm nl}}$. Evidently, the upper plots show a large scatter of the characteristic overdensity
of the halos throughout the different realizations, and a significant discrepancy in the total mass, as it clearly appears from
the large radii behavior of the density profiles in both plots. This confirms once again that for this type of simulations we are actually observing a superposition
of linear and nonlinear effects even when considering the most highly nonlinear objects in our simulation box.
It is then true, as noticed by \citet{Li_Barrow_2010b}, that the suppression of the EDE component in the Hubble function (cyan line)
and of the mass decrease of CDM particles (dotted orange line) determine the most significant increase (even with respect to the velocity-dependent acceleration) 
of the halo overdensity in the inner regions, but the same increase is visible at all radii, as it clearly appears when looking at the behavior of these models in the outskirts of the halos.
This however does not happen for the case where the velocity-dependent term is switched off (blue line), which shows a comparable characteristic overdensity
at the virial radius to the $\Lambda $CDM model (black line) and to the full interacting DE simulation (red line).
This superposition of linear and nonlinear effects can then lead to the wrong conclusion -- claimed as a partial discrepancy with the early results of \citet{Baldi_etal_2010} --
 that the background evolution and the mass
variation of CDM particles are the most important features of interacting DE models in the nonlinear regime.

This is clearly not true if we now consider the two bottom panels of Fig.~\ref{fig:profiles}, where the same halos are compared 
for the simulations where the individual effects are swhitched off only at low redshift. As we already stressed above, this produces
halos with a much more similar environment and with a comparable formation redshift, and allows a much clearer identification
of corresponding halos in the simulation set. In fact, from the bottom panels of Fig.~\ref{fig:profiles} one can notice that the halos in
all the simulations have the same characteristic overdensity at the virial radius, nevertheless showing very interesting departures from 
the $\Lambda $CDM and from the full interacing DE simulation in the inner regions. 
Therefore, this normalization allows to test the nonlinear behavior in the halo core in an almost independent way from the linear normalization of the surrounding environment.

Using this more meaningful procedure for the comparison of the different effects of the DE-CDM interaction shows results
in full agreement with what found by \citet{Baldi_etal_2010}. The velocity-dependent acceleration (blue line) is clearly the most important effect
in lowering the inner overdensity of massive halos, such that its suppression determines the largest overdensity increase in the halo cores. 
As already found by \citet{Baldi_etal_2010}, the mass variation has a minor but still significant effect, while the fifth force shows practically
no impact on the internal dynamics of the halos.

These results therefore fully address the apparent discrepancy claimed by \citet{Li_Barrow_2010b} that found evidence
for the opposite hierarchy of the velocity-dependent acceleration and mass variation effects in lowering halos inner overdensities.
The discrepancy only arises as a consequence of the numerical procedures adopted in the study of \citet{Li_Barrow_2010b} that 
determines a superposition of linear and nonlinear effects of the velocity-dependent acceleration, and a superposition of ``mass constancy" (\ie $\dot{m}_{c}=0$)
and ``mass normalization" (\ie the effective $\Omega _{c}$) for what concerns the variable mass of CDM particles.

\section{Conclusions}
\label{concl}

Interacting DE models have become very popular in recent years as a viable alternative to the standard $\Lambda $CDM cosmological model.
The investigation of these alternative cosmologies has shown significant progresses by moving from the study of the background evolution
to the analysis of linear perturbations effects and ultimately to the impact of interacting DE models on nonlinear structure formation
by means of specific modifications of N-body algorithms.

The main effects through which an interaction between DE and CDM can affect the growth of density perturbations range from
a modified background expansion, to a time variation of the CDM particle mass, to a long-range attractive ``fifth force" between
CDM particles and a ``modified inertia" in the form of a new velocity-dependent acceleration. 

In this work we have presented a detailed study of the impact that interacting DE models have on linear and nonlinear
structure formation processes. Our work significantly extends previous analyses and aims at a direct comparison between
different numerical procedures previously adopted in the literature to investigate the relative importance of each of the above mentioned physical
effects that characterize interacting DE cosmologies. 

By means of a suitable modification of the N-body code {\small GADGET-2}, we have performed a series of collisionless cosmological N-body
simulations for a standard $\Lambda $CDM model and for an interacting DE model with coupling $\beta =0.24$.
Such a large value for the coupling is already ruled out by several observational probes, however our aim here
was not to describe an observationally viable scenario but rather to explore in detail how an interaction between DE and CDM
affects structure formation, and a large coupling clearly makes this task more easily achievable by amplifying the effects under investigation.

In addition to the $\Lambda $CDM and the interacting DE simulations, we have carried out other 8 simulations in which each 
of the specific effects of the DE-CDM interaction has been in turn artificially suppressed, in order to quantify its relative contribution
to the different peculiar features of interacting DE cosmologies. In doing so, we have applied two different procedures previously 
adopted in the literature, namely the selective suppression of each individual effect only at the latest stages of structure formation
when most of the nonlinear processes take place \citep[as first proposed in][]{Baldi_etal_2010}, and the switch off of the different effects
during the whole simulations \citep[as more recently done by][]{Li_Barrow_2010b}. Our analysis therefore allows a direct comparison
of these two methods and provides a direct way to test the relative importance of the different features of interacting DE models at high 
and low redshifts.
\ \\

As a first test we have studied the impact of each individual effect of the DE-CDM interaction on the evolution of the matter power spectrum at different
redshifts. The global effect of interacting DE on the matter power spectrum is to suppress power at small scales while the
large scale amplitude has the same normalization as for $\Lambda $CDM at $z=0$.
This first test already shows a significant difference between the two numerical procedures mentioned above. In fact, the suppression
of any of the effects of the interaction right from the start of the simulations determines a different linear growth of density perturbations that
produces a significant scatter in the large scale normalization of the power spectra at $z=0$. As a consequence, the final power spectrum of each of these simulations
is modified at small scales by a superposition of the linear and nonlinear impact of the suppressed effect, making a direct test
of the relative importance of each effect quite difficult. 

On the contrary, if the suppression is limited to low redshifts, the large scale normalization
of the power spectrum remains consistent with the original $\Lambda $CDM and interacting DE models, and the small scale differences from 
simulation to simulation will be due only to the nonlinear impact of each specific effect. With this more suitable comparison procedure we have clearly shown
that the velocity-dependent acceleration of CDM particles is the most important effect in suppressing small scale power, with the mass variation
playing a minor but still significant role, while the fifth force has basically no effect in the nonlinear regime.
\ \\

We have also tested how each of these effects alter the relative evolution of density perturbations in the uncoupled baryonic component
as compared to CDM. In this case, the hierarchy of the different effects is reversed with respect to the impact on the power spectrum, and the 
fifth force shows the most important contribution to the faster growth of CDM perturbations with respect to baryonic perturbations both
at linear and nonlinear scales. This different growth gives rise to the so called ``gravitational bias" between the two components which 
determines a significant baryon depletion of collapsed objects as compared to $\Lambda $CDM.
Therefore, the fifth force is found to be the driving mechanism for the reduced baryon fraction of massive halos that characterize interacting DE cosmologies.
\ \\

We have then investigated the impact on the halo mass function of each of the effects under study, where once again the two different numerical 
procedures used in our analysis show significant differences. Also in this case, suppressing individual effects from the beginning of the simulation
determines a large scatter in the statistical distribution of collapsed objects at $z=0$ with respect to the original $\Lambda $CDM and full interacting
DE simulations that by construction give very similar mass functions at the present time.  This scatter is maximum at large masses
and shows how, in the simulations run with this procedure, structures cannot be expected to form in the same locations and have similar
masses and formation histories from simulation to simulation. This clearly makes any attempt to directly compare individual halos in different simulations
particularly hard, since it will be difficult to safely identify objects in different runs as being the same structure, as we have directly shown in our
analysis. 

On the contrary, again, we showed that suppressing individual effects of the DE-CDM interaction only at the latest stages of structure formation
produces very similar statistical distributions of halos at the present time, with basically no scatter around the original mass functions of $\Lambda $CDM
and of the full interacting DE model. This procedure therefore results clearly more suitable also in order to perform direct comparisons between individual halos in
different numerical realizations, as the spatial locations of bound objects as well as their local environment and formation histories will be very similar from simulation
to simulation.
\ \\

We have finally studied the relative importance of each individual effect of the DE-CDM interaction on the radial density profiles of CDM halos.
Consistently with the results found for the halo mass functions, we have shown that a suppression of any specific effect from high redshifts
determines a large scatter of the characteristic overdensity of the halos at $z=0$, which have a very different total mass and show significant
differences in the amplitude of the density profiles at all radii. This behavior again witnesses the superposition of linear and nonlinear effects 
when this numerical procedure is adopted. 

On the other hand, switching off the individual effects only at low redshifts preserves the characteristic 
overdensity of corresponding halos that show the same amplitude of their density profiles at large radii. As a consequence, the impact
of each individual effect on the internal dynamics and on the distortion of the profile at small radii is much more clearly visible and can be safely
disentangled from the linear normalization of the environment in which each halo is embedded.
By adopting this more suitable procedure, we found once again that the most relevant effect in reducing the inner overdensity of halos
in interacting DE models is given by the velocity-dependent acceleration of CDM particles, whereas the mass variation has a minor but 
not negligible impact, while the fifth force shows no influence whatsoever in this highly nonlinear context.

These results are in full agreement with what previously found by \citet{Baldi_etal_2010} and show that the discrepancies claimed by
\citet{Li_Barrow_2010b} are essentially due to the different procedure adopted in their numerical study, which determines a 
superposition of linear and nonlinear effects and is therefore less suitable to compare interacting DE models in the highly nonlinear
regime of structure formation.
\ \\

To conclude, we have performed a wide and detailed study of interacting DE cosmologies aimed at comparing
the relative importance that each of the new physical effects that characterize these models has in 
affecting linear and nonlinear structure formation processes. Our study significantly extends previous works and allows a direct
comparison between different numerical procedures recently adopted in the literature.
Our outcomes fully confirm the early results of \citet{Baldi_etal_2010}, and show that the apparent discrepancies
found by \citet{Li_Barrow_2010b} are actually due to the different numerical setup adopted by the latter work, which 
inevitably determines a superposition of linear and nonlinear effects and
is therefore not particularly suitable to compare the different characteristic 
features of interacting DE models in the nonlinear regime.

\section*{Acknowledgments}

This work has been supported by 
the DFG Cluster of Excellence ``Origin and Structure of the Universe''.
All the numerical simulations have been performed on the OPA cluster at the RZG computing centre in Garching.

\bibliographystyle{mnras}
\bibliography{baldi_bibliography}

\label{lastpage}

\end{document}